\documentclass[twoside,preprintnumbers,amsmath,amssymb,pacs,shownopacs,nofootinbib]{revtex4-2}
\usepackage{graphicx,subfigure}
\usepackage{braket,euscript,enumitem}
\usepackage{fancyvrb}

\usepackage[Gray,squaren,thinqspace,thinspace]{SIunits}

\usepackage{xcolor}
\usepackage{bbm}
\usepackage{dsfont}

\newcommand{\MSbar}{\overline{\mbox{MS}}}

\newcommand{\lms}{\Lambda_{\overline{\mbox{\tiny{MS}}}}}
\newcommand{\mx}{\ensuremath{\mathsf}}
\newcommand{\Res}[1][]{\ensuremath{\underset{#1}{\mathcal Res}}}
\newcommand{\tr}{{\rm tr}\,}
\allowdisplaybreaks[2]
\newcommand{\citeapp}{Appendix~\ref{backgroundren}}

\newcommand{\citeappdrie}{Appendix~\ref{sums}}
\newcommand{\citeappvier}{Appendix~\ref{NL}}

\begin{document}

\title{{\bf Polyakov loop, gluon mass, gluon condensate and its asymmetry near deconfinement}}

\author{D.~Dudal$^{1,2}$,  D.M.~van Egmond$^{3}$, U.~Reinosa$^{3}$, D.~Vercauteren$^{4,5}$}
\email{david.dudal@kuleuven.be, duifje.van-egmond@polytechnique.edu, urko.reinosa@polytechnique.edu, vercauterendavid@duytan.edu.vn}
\affiliation{$^1$ KU Leuven Campus Kortrijk--Kulak, Department of Physics, Etienne Sabbelaan 53 bus 7657, 8500 Kortrijk, Belgium\\
$^2$ Ghent University, Department of Physics and Astronomy, Krijgslaan 281-S9, 9000 Gent, Belgium \\
$^3$ Centre de Physique Th\'{e}orique, CNRS,
Ecole Polytechnique, IP Paris, F-91128, Palaiseau, France \\
$^4$ Institute of Research and Development, Duy Tan University, Da Nang 550000, Vietnam \\
$^5$ Faculty of Natural Sciences, Duy Tan University, Da Nang 550000, Vietnam}


\begin{abstract}
We consider a BRST invariant generalization of the ``massive background Landau gauge'', resembling the original Curci--Ferrari model that saw a revived interest due to its phenomenological success in modeling infrared Yang--Mills dynamics, including that of the phase transition. Unlike the Curci--Ferrari model, however, the mass parameter is no longer a phenomenological input but it enters as a result of dimensional transmutation via a BRST invariant dimension two gluon condensate. The associated renormalization constant is dealt with using Zimmermann's reduction of constants program which fixes the value of the mass parameter to values close to those obtained within the Curci--Ferrari approach. Using a self-consistent background field, we can include the Polyakov loop and probe the deconfinement transition, including its interplay with the condensate and its electric--magnetic asymmetry. We report a continuous phase transition at {$T_c\approx 0.230$~GeV} in the SU(2) case and a first order one at {$T_c\approx 0.164$~GeV}  in the SU(3) case, values which are again rather close to those obtained within the Curci--Ferrari model at one-loop order.
\end{abstract}
\maketitle

\section{Introduction}
It is well accepted from non-perturbative Monte Carlo lattice simulations that $SU(N)$ Yang--Mills gauge theories in the absence of fundamental matter fields undergo a deconfining phase transition at a certain critical temperature \cite{Lucini:2003zr,Lucini:2012gg}. This transition corresponds to the breaking of a global $\mathbb{Z}_N$ center symmetry when the Euclidean temporal direction is compactified on a circle, with circumference proportional to the inverse temperature, see e.g.~\cite{Svetitsky:1985ye,Greensite:2003bk}. The vacuum expectation value of the Polyakov loop \cite{Polyakov:1978vu} serves as an order parameter for this symmetry, and has as such inspired an ongoing research activity into its dynamics, see e.g.~\cite{Fukushima:2003fw,Schaefer:2007pw,Maas:2011ez,Fischer:2012vc}.
Even in the presence of dynamical quark degrees of freedom, in which case the center symmetry is broken explicitly, the Polyakov loop remains the best observable to capture the cross-over transition, see  \cite{Borsanyi:2010bp,Bazavov:2011nk} for ruling lattice QCD estimates. Since the transition temperature is of the order of the scale at which the considered gauge theories, including QCD, become strongly coupled, it is a highly challenging endeavour to get reliable estimates for the Polyakov loop correlators, including its vacuum expectation value, analytically. This is further complicated by the non-local nature of the loop. These features highlight the sheer importance of lattice gauge theories to allow for a fully non-perturbative computational framework. Nonetheless, analytical takes are still desirable to offer a complementary view at the same physics, in particular as lattice simulations do also face difficulties when the physically relevant small quark mass limit must be taken, next to the issue of potentially catastrophic sign oscillations at finite density \cite{Fukushima:2010bq,deForcrand:2002hgr}.

Over the last two decades, a tremendous effort has been put into the development and application of Functional Methods to QCD, including the respective hierarchies of Dyson--Schwinger and Functional Renormalization Group equations \cite{vonSmekal:1997ohs, Alkofer:2000wg, Zwanziger:2001kw, Fischer:2003rp, Bloch:2003yu, Aguilar:2004sw, Boucaud:2006if, Aguilar:2007ie, Boucaud:2008ky, Fischer:2008uz, Rodriguez-Quintero:2010qad,Wetterich:1992yh, Berges:2000ew, Pawlowski:2003hq, Fischer:2004uk,Pawlowski:2005xe,Cyrol:2016tym,Dupuis:2020fhh} as well a variational approaches based on the Hamiltonian formulation or on $N$-particle-irreducible effective actions \cite{Schleifenbaum:2006bq,Quandt:2013wna,Quandt:2015aaa,Carrington:2007ea,Alkofer:2008tt,York:2012ib,Fister:2013bh}. These methods are quite successful in describing vacuum properties of the theory as well as finite temperature/density aspects. They all rely, in one way or another, on the decoupling behavior of gluons in the Landau gauge, as dictated by results from lattice simulations \cite{Cucchieri:2007rg, Cucchieri:2008fc, Bornyakov:2008yx, Cucchieri:2009zt, Bogolubsky:2009dc, Bornyakov:2009ug, Dudal:2010tf,Duarte:2016iko}. More recently, a more phenomenological approach has been put forward based on the use of the Curci--Ferrari model \cite{Tissier:2010ts,Tissier:2011ey,Pelaez:2021tpq}. The rationale behind the latter is that the standard Faddeev--Popov Landau gauge action, although well grounded in the ultraviolet, is incomplete in the infrared due to the presence of Gribov copies, and, therefore, needs to be extended. The hypothesis put forward in \cite{Tissier:2010ts} is that a dominant contribution to this (to date unknown) gauge-fixed action is provided by a gluon mass term, which relates to the decoupling behavior of the Landau gauge gluon propagator on the lattice. One of the attractive features of the Curci--Ferrari model is that it is perturbative in nature, at least in its applications to pure Yang--Mills theories. In fact, with just one additional parameter to adjust, it has allowed one to retrieve many of the Euclidean properties of these theories in the vacuum and at finite temperature \cite{Pelaez:2021tpq}. In its applications to QCD, the perturbative nature of the pure glue sector allows one, in combination with an expansion in the inverse number of colors, to devise a systematic expansion scheme controlled by two small parameters and whose first orders are computionally tractable \cite{Pelaez:2017bhh,Pelaez:2020ups}.

The surprising ability of the Curci--Ferrari model in reproducing well known properties of pure YM theories has lead to the question of whether it could be derived (in its present form or with some amendments) from a proper account of the Gribov copies \cite{Serreau:2012cg,Tissier:2017fqf,Reinosa:2020skx}. Here we would like to investigate another possibility following the work of \cite{Verschelde:2001ia} and based on the dynamical generation of dimension two gluon condensates within the strict Faddeev--Popov set-up. The idea here is that upon generation of a dynamical gluon mass, the Gribov copies will be accounted for, at least partially. So, more than a consequence of taking into account the Gribov copies, the gluon mass will here appear as a self-generated cure for the Gribov problem within the Faddeev--Popov framework. In what follows we would like to investigate these ideas, in particular how they allow one to describe salient features of YM theory such as the deconfinement transition. Since the Curci--Ferrari model has taught us that, once a mass is generated, certain features become accessible to perturbation theory, we shall consider a simple one-loop calculation as a start.

Because the decoupling behavior as observed on the lattice extends beyond the Landau gauge to linear covariant gauges \cite{Bicudo:2015rma,Napetschnig:2021ria}, it will be important to make sure that the dynamical mass generation mechanism applies independently of the gauge-fixing parameter. Moreover, for the dynamically generated mass to carry a physical significance, it should be associated to a BRST invariant gluon condensate. A central notion to achieve this is that of a BRST invariant gluon field  \cite{Capri:2016ovw,Capri:2016gut} which we discuss in Sec.~\ref{sec:BRST_gluon}, together with its extension in the presence of a background gauge field, required for the study of the Polyakov loop. We also show that the BRST-invariant gluon field can be replaced by the original gluon field in the limit of a vanishing gauge-fixing parameter, which will later facilitate the computations. In Sec.~\ref{LCO}, we introduce the BRST invariant dimension two gluon condensate, together with its BRST invariant asymmetry at finite temperature. This asymmetry was proposed in past Landau gauge-fixed lattice QCD work \cite{Chernodub:2008kf} to constitute yet another probe of the deconfinement transition. More generally, we expect an interesting interplay between {the condensate, and thus the mass,} and the Polyakov loop at finite temperature. In Sec.~\ref{sec:pot}, we evaluate the effective potential for the background field (related to the Polyakov loop), the BRST-invariant condensate (related to the mass) and the asymmetry in this BRST invariant condensate. Our results for the three observables across the deconfinement transition are gathered in Sec.~\ref{sec:results} together with a discussion relating to the Curci--Ferrari model.

 \section{BRST invariant gluon field $A^h$}\label{sec:BRST_gluon}

To set the stage, we will first briefly introduce our construction at zero temperature and without background gauge fields, summarizing a larger paper in preparation \cite{prep} based on earlier work \cite{Verschelde:2001ia}, before extending it in the presence of the Polyakov loop via the background field method.

\subsection{The case of linear covariant gauges} \label{Ah}
We start from the Yang--Mills action in a linear covariant {(LC)} gauge and in $d$ Euclidean space dimensions:
\begin{equation}\label{eq:LC}
	{S_\text{LC}} = \int d^dx \left(\tfrac14 F_{\mu\nu}^a F_{\mu\nu}^a + \tfrac\alpha2 b^a b^a + ib^a \partial_\mu A_\mu^a + \bar c^a \partial_\mu D_\mu^{ab}c^b\right) \;,
\end{equation}
where $c$ and $\bar c$ are the ghost and anti-ghost fields, $b$ is the Nakanishi--Lautrup field enforcing the gauge condition, and $\alpha$ is the gauge parameter. As we are eventually interested in the dimension two gluon condensate $\langle A_\mu^2\rangle$ while preserving BRST invariance, we need a BRST invariant version of the $A_\mu^a$ field. In order to construct this, we insert into the corresponding path integral the following unity \cite{Capri:2018ijg,prep}:
\begin{subequations}
\begin{gather}
	1 = \mathcal N \int [\mathcal D\xi\mathcal D\tau\mathcal D\bar\eta\mathcal D\eta] e^{-S_h} \;, \label{unity} \\
	S_h = \int d^dx \left(i \tau^a\partial_\mu(A^h)_\mu^a + \bar\eta^a \partial_\mu (D^h)_\mu^{ab}\eta^b\right) \;,
\end{gather}
where $\mathcal N$ is a normalization and $(D^h)_\mu^{ab}$ is the covariant derivative containing only the composite field $(A^h)_\mu^a$. This local but non-polynomial composite field object is defined as:
\begin{gather}
	(A^h)_\mu = h^\dagger A_\mu h + \tfrac ig h^\dagger \partial_\mu h \;, \label{3} \\
	h = e^{ig\xi} = e^{ig\xi^a T^a} \;,
\end{gather}
\end{subequations}
where the $T^a$ are the generators of the gauge group SU($N$). The $\xi^a$ are similar to Stueckelberg fields, while $\eta^a$ and $\bar\eta^a$  are additional (Grassmanian) ghost and anti-ghost fields. They serve to account for the Jacobian arising from the functional integration over $\tau^a$ to give a Dirac delta functional of the type $\delta(\partial_\mu(A^h)_\mu^a)$. That Jacobian is similar to the one of the Faddeev--Popov operator, and is supposed to be positive which amounts to removing a large class of infinitesimal Gribov copies, see \cite{Capri:2015ixa}. Here, positivity can be checked a posteriori by means of the ghost propagator, the (expectation value of the) inverse Faddeev--Popov operator.  Notice that in mere perturbation theory, this is not the case, but the dynamical mass to be discussed will be large enough to ensure it dynamically, leading to a kind of  ``self-cured'' Gribov ambiguity. This is a different approach than the Gribov--Zwanziger one, in which case positivity is imposed a priori \cite{Gribov:1977wm,Zwanziger:1989mf}, albeit with similar end result. In fact, this strategy of having a positive ghost propagator is also the one employed in e.g.~the functional Dyson--Schwinger approach \cite{Fischer:2008uz}.\footnote{Related to this dicussion, we note that (\ref{unity}) is a priori an approximation since it ignores the Gribov copies. In the presence of a dynamically generated mass, the contribution of some of these copies is suppressed, in particular those outside the Gribov region, in a fashion similar (but not equivalent) to the Gribov--Zwanziger approach.}

Expanding \eqref{3}, one finds an infinite series of local terms:
\begin{equation}\label{reeks}
	(A^h)_\mu^a = A_\mu^a - \partial_\mu\xi^a - gf^{abc}A_\mu^b\xi^c - \tfrac g2 f^{abc}\xi^b\partial_\mu\xi^c + \cdots \;.
\end{equation}
The unity \eqref{unity} can be used to stay within a local setup for an on-shell non-local quantity $(A^h)_\mu^a$ that can be added to the action. Notice that the multiplier $\tau^a$ implements $\partial_\mu(A^h)_\mu^a = 0$ which, when solved iteratively for $\xi^a$
\begin{subequations}
\begin{equation}
	\xi_* = \frac1{\partial^2} \partial_\mu A_\mu + ig \frac1{\partial^2} \left[\partial_\mu A_\mu,\frac1{\partial^2} \partial_\nu A_\nu\right] + \cdots \;,
\end{equation}
gives the (transversal) on-shell expression
\begin{gather}
	(A^h)_\mu = \left(\delta_{\mu\nu} - \frac{\partial_\mu\partial_\nu}{\partial^2} \right) \left(A_\nu + ig \left[A_\nu,\frac1{\partial^2} \partial_\lambda A_\lambda\right] + \cdots\right) \;,
\end{gather}
\end{subequations}
clearly showing the non-localities in terms of the inverse Laplacian. One can see that $A^h \to A$ when $A_\mu^a$ is in the Landau gauge $\partial_\mu A_\mu^a = 0$. We refer to \emph{e.g.} \cite{DellAntonio:1991mms,Lavelle:1995ty,Capri:2015ixa,Capri:2018ijg,prep} for more details. It can be shown that $A^h$ is gauge invariant order per order, which is sufficient to establish BRST invariance. We will have nothing to say about large gauge transformations.

Mark that $(A^h)_\mu^a$ is formally the value of $A_\mu^a$ that (absolutely) minimizes the functional
\begin{equation}\label{A2}
	\int d^dx A_\mu^a A_\mu^a
\end{equation}
under (infinitesimal) gauge transformations $\delta A_\mu^a = D_\mu^{ab} \omega^b$, see e.g.~\cite{DellAntonio:1991mms,Lavelle:1995ty,Capri:2015ixa}. As such,
\begin{equation}\label{nummer}
\int d^dx {(A^h)}_\mu^a {(A^h)}_\mu^a=\min_{{\mbox{\tiny{gauge orbit}}}} \int d^d x A_\mu^a A_\mu^a\,,
\end{equation}
In practice, we are only (locally) minimizing the functional via a power series expansion \eqref{reeks} coming from infinitesimal gauge variations around the original gauge field $A_\mu^a$, whereas the extremum being a minimum is accounted for if the Faddeev--Popov operator (second order variation that is) is positive. This is related to the Gribov copy problem and will be ignored here in the definition of our ${(A^h)}_\mu^a$ or the unity. We will come back to why this is a posteriori allowed.

We will later on generalize this construction in the presence of a background gauge field, including the proof that, for expectation values of gauge invariant operators, the non-local $A_\mu^h$ can be replaced by the local $A_\mu$ when using the Landau gauge,  corresponding to the $\alpha\to0$ case of the linear covariant gauges. The positivity of the Faddeev--Popov operator will also play a role here. But summarizing, at the level of expectation values of gauge invariant operators, {the original action (\ref{eq:LC})} and the one given by
\begin{eqnarray}\label{nieuwactie}
{S_\text{LC}+S_h=}\int d^dx \left(\frac{1}{4} F_{\mu\nu}^2+\frac{\alpha}{2}b^2\!+ib^{a}\partial_{\mu}A^{a}_{\mu}
+\bar{c}^{a}\partial_{\mu}D^{ab}_{\mu}c^{b}+i\tau^a\partial_\mu(A^h)_\mu^a + \bar\eta^a \partial_\mu (D^h)_\mu^{ab}\eta^a\right)
\end{eqnarray}
are perturbatively fully equivalent. The renormalizability analysis for generic $\alpha$ can be found in e.g.~\cite{Capri:2016ovw}. For completeness, the BRST invariance is generated by the operator $s$ defined as
\begin{equation} \label{ourbrst2}
	sA_\mu^a = -D_\mu^{ab}c^b \;, \qquad sc^a = \tfrac12 gf^{abc}c^bc^c \;, \qquad s\bar c^a = -ib^a \;,
\end{equation}
and all other transformations zero.

\subsection{Including the Polyakov loop}
Our aim is to investigate the confinement/deconfinement phase transition {of Yang-Mills theory.} The standard way to achieve this goal is by probing the Polyakov loop order parameter,
\begin{equation}
	\mathcal{P} = \frac{1}{N}\tr \Braket{P e^{ig\int_{0}^{\beta}dt \ A_{0}(t,x)}} \;,
\end{equation}
with $P$ denoting path ordering, needed in the non-Abelian case to ensure the gauge invariance of $\mathcal{P}$. In analytical studies of the phase transition involving the
Polyakov loop, one usually imposes the so-called ``Polyakov gauge'' on the gauge field, in which case the time-component $A_{0}$ becomes diagonal and independent of (imaginary) time{: $\langle A_{\mu}(x)\rangle = \langle A_{0}\rangle \delta_{\mu 0}$, with $\langle A_{0}\rangle$ belonging to the Cartan subalgebra of the gauge group. In the SU(2) case for instance, the Cartan subalgebra is one-dimensional and can be chosen to be generated by $t^3\equiv\sigma^3/2$, so that $\langle A^{a}_{0}\rangle = \delta^{a3}\langle A^3_0\rangle \equiv \delta^{a3} \langle A_0\rangle$.} More details on Polyakov gauge can be found in \cite{Marhauser:2008fz,Fukushima:2003fw,Ratti:2005jh}. Besides the trivial simplification of the Polyakov loop, when imposing the Polyakov gauge it turns out that the quantity $\Braket{A_{0}}$ becomes a good alternative choice for the order parameter instead of $\mathcal{P}$, see \cite{Marhauser:2008fz} for an argument using Jensen's inequality for convex functions, see also \cite{Braun:2007bx,Reinhardt:2012qe,Reinhardt:2013iia}. For other arguments based on the use of Weyl chambers and within other gauges (see below), see \cite{Reinosa:2015gxn,Herbst:2015ona,Reinosa:2020mnx}.

{As explained in  \cite{Braun:2007bx,Marhauser:2008fz,Reinosa:2014ooa}, in the SU(2) case at leading order we then simply find, using the properties of the Pauli matrices,
\begin{equation}
	\mathcal{P}=\cos\frac{r}{2}\;,
\end{equation}
where we defined
\begin{equation}
  r=g\beta \langle A_0\rangle\;,
\end{equation}
with $\beta$ the inverse temperature. This way, $r=\pi$ corresponds to the ``unbroken symmetry phase'' (confined or disordered phase), equivalent to $\Braket{{\cal P}} = 0$; while $0<r<\pi$ corresponds to the ``broken symmetry phase'' (deconfined or ordered phase), equivalent to $\Braket{{\cal P}} \neq 0$. Since $\mathcal{P}\propto e^{-F/T}$ with $T$ the temperature and $F$ the free energy of a heavy quark, it becomes clear that in the unbroken phase (because { it} is in the unbroken phase where the center symmetry is manifest: $\Braket{{\cal P}} = 0$), an infinite amount of energy would be required to actually get a free quark. The broken/restored symmetry referred to is the $\mathbb{Z}_N$ center symmetry of a pure gauge theory (no dynamical matter in the fundamental representation). With a slight abuse of language, we will refer to the quantity $r$ as the Polyakov loop hereafter.}

It is however a highly non-trivial job to actually compute $r$. An interesting way around was worked out in  \cite{Braun:2007bx,Marhauser:2008fz,Reinosa:2014ooa}, where it was shown that similar considerations apply within Landau--DeWitt gauges, a generalization of the Landau gauge in the presence of a background (see the next section for more details). The background needs to be seen as a field of gauge-fixing parameters and, as such, can be chosen at will a priori. However, specific choices turn out to be computationally more tractable while allowing one to unveil more easily the center-symmetry breaking mechanism. In particular, for the particular choice of {\it self-consistent backgrounds} which are designed to coincide with the thermal gluon average at each temperature, it could be shown that the background becomes an order parameter for center-symmetry as it derives from a center-symmetric background effective potential (see below).

Moreover, non-perturbative physics was parameterized by a phenomenological mass parameter, akin to using a Curci--Ferrari version of the background Landau gauge \cite{Reinosa:2014ooa,Reinosa:2014zta}. This was based on earlier successful attempts to model $T=0$ Yang--Mills propagators and vertices, see \cite{Tissier:2010ts,Tissier:2011ey} for the initial works, and \cite{Pelaez:2021tpq} for a recent overview. The Curci--Ferrari mass was fixed from a dedicated fit to zero temperature lattice gluon and ghost propagator data in absence of a background, see e.g.~\cite{Gracey:2019xom}, but despite its nice consequences and quite good results compared to other non-perturbative approaches, it remains a bit uncomfortable that one needs to introduce a mass scale by hand. If we could recover a dynamical gluon mass from a first principles setup, this would reduce the dependence on external parameters or input. Of course, this does not necessarily entail we will end up with the exact Curci--Ferrari model or the background version of \cite{Reinosa:2014ooa}, but this is evidently of no concern, since the Curci--Ferrari was always supposed to be an effective way of modelling gauge fixing beyond standard perturbation theory.

That a proper mass scale can emerge from the Yang--Mills dynamics can already be appreciated from earlier works like \cite{Verschelde:2001ia}, based on the introduction of the non-local but gauge invariant gluon condensate $\Braket{A^2}_{\min}$, which reduces to $\Braket{A^2}$ in the Landau gauge in \cite{Gubarev:2000eu,Gubarev:2000nz,Boucaud:2000ey,Boucaud:2001st}. In fact, $\Braket{A^2}_{\min}=\Braket{A^h A^h}$, see the discussion below \eqref{A2}.

Other approaches in which (dynamical) gluon mass scales played a role are, for example, \cite{Aguilar:2004sw,Aguilar:2006gr,Dudal:2008sp,Aguilar:2008xm,Fischer:2008uz,Aguilar:2011ux,Dudal:2011gd,Cyrol:2016tym,Capri:2016aif,Capri:2017npq,Cyrol:2017ewj,Comitini:2017zfp,Siringo:2018uho,Horak:2022aqx,prep}.

\subsection{BRST invariant gluon field in presence of a background}\label{ahsectie}
We are thus ultimately interested in investigating the spontaneous generation of a gluon mass. In the presence of a background and in the Landau--DeWitt gauge, renormalization (see \citeapp) imposes that this mass (and an asymmetry in this mass, for which see Section \ref{asym}) should couple only to the quantum fields, \emph{i.e.}~the full field minus the background value. This is because quantum fields and background renormalize differently.

In order to implement this, the formalism of Section \ref{Ah} needs to be slightly adapted. Assume a background $\bar A_\mu^a$ such that the full gluon field $a_\mu^a$ can be written as
\begin{equation}
	a_\mu^a = \bar A_\mu^a + A_\mu^a \;,
\end{equation}
where $A_\mu^a$ now denotes the quantum part only. As there is a background, it is convenient to use the Landau--DeWitt (LDW) gauge fixing condition or Landau background gauge
\begin{equation} \label{ldwcond}
	\bar D_\mu^{ab} (a_\mu^b-\bar A_\mu^b) = 0 \;,
\end{equation}
where $\bar{D}^{ab}_{\mu} =\delta^{ab}\partial_{\mu} - gf^{abc}\bar{A}^{c}_{\mu}$ is the background covariant derivative. The Landau--DeWitt gauge can be defined as corresponding to the (local) minima of the functional
\begin{equation}\label{functionaal2}
	\int d^dx (a_\mu^a-\bar A_\mu^a)^2 = \int d^dx A_\mu^a A_\mu^a
\end{equation}
under infinitesimal gauge transformations $\delta a_\mu^a = \delta A_\mu^a = D_\mu^{ab} \omega^b$. We refer to \cite{Zwanziger:1982na,Cucchieri:2012ii} for more details. Also here, the extremum will be a minimum upon having a positive Faddeev--Popov operator $\bar D_\mu^{ab} D_\mu^{bc}$.

Mark that the background does not transform here, the entire gauge transformation is associated to the quantum part. In principe, the gauge transformation can be distributed over the quantum and classical part, but the choice we make is the most natural one and relates best to the BRST operator to be introduced later (see \eqref{ourbrst} and \citeapp), at vanishing external sources. The BRST operator then also leaves the background field untouched.

Mark further that invariance under gauge transformations of the background (under which the quantum part transforms as a matter field) is a separate issue and is not a problem in our case (unlike in \cite{Dudal:2017jfw} for example), see subsection \ref{bgi} and \citeapp.

{Finally, we note that, similarly to the case in the absence of background, we can extend the Landau-deWitt gauge into to a linear covariant version of it, which we refer to as linear background covariant gauge
\begin{equation}\label{nieuweactie2bis}
S_{\mbox{{\tiny bLC}}} = \int d^4x \left(\frac14(F_{\mu\nu}^a)^2 + \tfrac\alpha2 b^a b^a + ib^a\bar D_\mu(a_\mu^a-\bar A_\mu^a) + \bar c^a\bar D_\mu D_\mu c^a\right).
\end{equation}}

We now need to construct the field $(a_\mu^a)^h$ obeying the Landau--DeWitt gauge
\begin{equation} \label{ldwcondh}
	\bar D_\mu^{ab} ((a_\mu^b)^h-\bar A_\mu^b) = 0 \;.
\end{equation}
In order to do this, we will perform an expansion in the quantum fields.\footnote{In this paper, we will always remain in the perturbative formalism.} In this paper, we only aim to do one-loop computations, such that first order in the quantum fields will suffice.

We write the necessary gauge transform as $\mathbbm1 + h_1 + \cdots$, where $h_1=ig \xi_1^a t^a$ is assumed first order in the quantum fields and the dots contain higher order term. Up to first order we have
\begin{equation}
	a_\mu^h = \bar A_\mu + A_\mu + \frac ig \bar D_\mu h_1 + \cdots \;,
\end{equation}
where we used that $h_1^\dagger = -h_1$ (which is a consequence of the unitarity of $\mathbbm1+h_1+\cdots$ at first order). Imposing the gauge condition \eqref{ldwcondh} yields
\begin{equation}
	\bar D_\mu A_\mu + \frac ig \bar D^2 h_1 + \cdots = 0 \qquad \Rightarrow \qquad \frac ig h_1 = -\frac1{\bar D^2} \bar D_\mu A_\mu \;.
\end{equation}
As such we get
\begin{equation} \label{ahexp}
	a_\mu^h - \bar A_\mu = \left(\delta_{\mu\nu} - \bar D_\mu \frac1{\bar D^2} \bar D_\nu\right) A_\nu + \cdots \;.
\end{equation}
At first order in the quantum fields, $(a_\mu^a)^h$ is indeed invariant under $\delta A_\mu^a = D_\mu^{ab} \omega^b = \bar D_\mu^{ab} \omega^b + \cdots$. After gauge-fixing with the Faddeev--Popov procedure, this will translate into invariance under .
\begin{equation} \label{ourbrst}
	sa_\mu^a = -D_\mu^{ab}c^b \;, \qquad sc^a = \tfrac12 gf^{abc}c^bc^c \;, \qquad s\bar c^a = -ib^a\;, \qquad s(\text{rest})=0 \;,
\end{equation}
which is actually the very same BRST operator as defined in \eqref{ourbrst2}, as $a_\mu$ is now the complete field.

Lastly, the steps leading to \eqref{nieuwactie} are now easily generalized. We can introduce a rather complicated unity
\begin{subequations}
\begin{gather}
	1 = \mathcal N \int [\mathcal D\xi\mathcal D\tau\mathcal D\bar\eta\mathcal D\eta] e^{-S_h} \;, \label{uniteit} \\
	S_h = \int d^dx \left(i\tau^a \bar D_\mu^{ab} (a_\mu^{h,b}-\bar A_{\mu}^b) + \bar\eta^a \bar D_\mu^{ab} D_\mu^{bc}[a^h]\eta^c\right) \;,
\end{gather}
\end{subequations}
{and replace action (\ref{nieuweactie2bis}) with}
\begin{equation}\label{nieuweactie2}
{S\equiv S_{\mbox{{\tiny bLC}}}+S_h=\int d^dx \left(\frac14(F_{\mu\nu}^a)^2 +\tfrac\alpha2 b^a b^a+ ib^a\bar D_\mu(a_\mu^a-\bar A_\mu^a) + \bar c^a\bar D_\mu D_\mu c^a+i \tau^a \bar D_\mu^{ab} (a_\mu^{h,b}-\bar A_{\mu}^b) + \bar\eta^a \bar D_\mu^{ab} D_\mu^{bc}[a^h]\eta^c\right) \,.}
\end{equation}
For the record, we refrain here from a  full-blown all-order algebraic analysis of the renormalizability of the theory defined by \eqref{nieuweactie2} and of the background mass operator (see the next subsection), this could be done by combining the technology of \citeapp, \cite{brstbackground} and \cite{Capri:2017npq}, even for a more general class of background gauge fixings as in \cite{Ferrari:2000yp}.

\subsection{ $a_\mu^h \to a_\mu$ in the path integral in the Landau--DeWitt gauge} \label{speciaal}
Analogously  to the $\bar A=0$ case, averages computed with either the action \eqref{nieuweactie2} or the original Yang--Mills one will not change anything at the level of physical observables, defined via the BRST cohomology at zero ghost charge\footnote{Which are the classically gauge invariant operators built from the gauge field, up to irrelevant BRST exact terms and terms with equation-of-motion contributions.}, as one is free to choose any gauge to work with in practice, and employing the Landau--DeWitt gauge, we may effectively replace $a_\mu^h$ with $a_\mu$.  A formal way to show that this substitution is valid for expectation values of gauge invariant operators $O_i(x_i)$ goes as follows. {Consider then the action $S$ as given in \eqref{nieuweactie2}, with $\alpha\to 0$, that is $S=S_{\mbox{\tiny LDW}}+S_h$.} As before, we will use $A_\mu:=a_\mu-\bar A_\mu$ to denote the quantum fluctuation, that is, the field integrated over. To avoid clutter in the notation, we will strip all fields of indices, and introduce the shorthand $\bar D D^h := \bar D_\mu^{ab} D_\mu^{bc}[a^h]$.  $\Phi$ collects all quantum fields.
\begin{equation}\label{stap1}
   \Braket{O_1(x_1)\ldots O_n(x_n)}_{S}= \frac{\int [\mathcal{D}\Phi] O_1(x_1)\ldots O_n(x_n)e^{-S}}{\int [\mathcal{D}\Phi] e^{-S}}\,.
\end{equation}
Mark that the gauge invariance of $O_i$ means that $O_i[a]=O_i[a^h]$. Integration over $b, \tau, \bar\eta, \eta$ leads to
\begin{equation}\label{stap2}
   \delta(\bar D A)\delta( \bar D(a^h-\bar A))\det(-\bar D D^h)\;.
\end{equation}
Using the perturbative solution $\xi_*$ of the constraint $\bar D(a^h-\bar A)=0$ (see Section~\ref{ahsectie} for the explicit solution at leading order), we may rewrite the second Dirac delta as
\begin{equation}\label{stap3}
   \delta( \bar D(a^h-\bar A))=\delta(\xi-\xi_*)\frac{1}{\left|\det \delta_\xi (-\bar D D^h)\right|_{\xi=\xi_*}}
\end{equation}
to facilitate the $\xi$-integration.

We also note that
\begin{equation}\label{stap4}
   \left.\delta_\xi\det (-\bar D D^h)\right|_{\xi=\xi_*}= \det(-\bar D D + \mathcal{O}( \bar D A))\;,
\end{equation}
where $\mathcal{O}(\bar D A)$ is a formal power series in $\bar D A$, starting at order $\bar D A$. We already used here that $\xi_*$ itself is also such power series. The other Dirac delta constraint then leads to a factor
\begin{equation}\label{stap5}
   \frac{\det (-\bar D D)}{\left|\det (-\bar D D)\right|}=1\;,
\end{equation}
so the integration over $\tau,\eta, \bar\eta$ effectively constitutes a unity and effectively replaces $a^h$ with $a$, at least if the last step is valid. This is the case if the Faddeev--Popov operator is positive, which is equivalent to stating that the second derivative of the functional \eqref{functionaal2} is positive, {\it i.e.}~that we end up in a (local) minimum. This positivity requirement is equivalent to removing infinitesimal Gribov gauge copies in the Landau--DeWitt gauge, which as we already discussed for the $\bar A=0$ case can be a posteriori checked by means of the ghost propagator and its positivity. As before, this is not the case when using perturbation theory around the perturbative vacuum, but it is the case when a sufficiently large dynamical mass is generated, see \cite{Reinosa:2016iml} for an explicit one-loop verification. For other work on Gribov copies in presence of a background, see for instance \cite{Zwanziger:1982na,Canfora:2015yia,Canfora:2016ngn,Dudal:2017jfw,Kroff:2018ncl,Junqueira:2020whg,Justo:2022vwa,Justo:2022vwa}.

Returning to the discussion below \eqref{nummer}, if a posteriori a sufficiently large dynamical mass is generated, the Gribov problem is thus (partially) tamed\footnote{We have nothing to say about ``large'' gauge copies.}, the a priori assumption of ignoring the Gribov problem in defining the (unique) perturbative series solution in the minimization of \eqref{functionaal2} makes sense, and this with or without background field $\bar A$. This also makes \eqref{stap3} valid, as otherwise we would need to include all possible solutions here.

Needless to say, the quantum effective potential of a BRST invariant operator is an example where the substitution $a_\mu^h\to a_\mu$ applies. Notice that in combination with the results of \citeapp, this also implies that the effective action for BRST invariant operators derived from the action \eqref{nieuweactie2} will enjoy the background gauge invariance, as follows from the Ward identity \eqref{backgroundwarid}. This gives an exact argument, complementing the one already provided below \eqref{achter}.

\section{BRST-invariant mass and asymmetry} \label{LCO}
This section presents a short review of the Local Composite Operator (LCO) formalism as proposed in \cite{Verschelde:2001ia} modified in the presence of a background field. The case without background is a special case hereof, and will be discussed in greater detail in \cite{prep}.

\subsection{$d=2$ gluon condensate}
As we want to work with a background field, it is more appropriate to use the Landau background gauge \cite{Abbott:1981ke} $\bar D_\mu^{ab} (A_\mu^b)^h = 0$ instead of the usual Landau gauge prescription. For other works in the Landau background gauge, see for example \cite{Binosi:2009qm,Binosi:2011ar,Binosi:2012pd,Gies:2022mar}.

A BRST analysis (for BRST in the background gauge, see for example \cite{Ferrari:2000yp,brstbackground}\footnote{In \cite{brstbackground}, the BRST transform of the background is nonzero: $s\bar A_\mu^a = \Omega_\mu^a$, where $\Omega_\mu^a$ is a ghost source. This source greatly simplifies the proof of renormalizability. The physical case, however, is recovered when $\Omega_\mu^a\to0$, with BRST variation \eqref{ourbrst}, under which $(a^h)_\mu^a$ is invariant.}) shows that, for the LCO formalism to stay renormalizable, the dimension-two operator to be used is
\begin{equation}
(a_\mu^h - \bar A_\mu)^2 \;.
\end{equation}
First, the source terms
\begin{equation} \label{lcobg}
\int d^dx \left(\tfrac12 J (a_\mu^h - \bar A_\mu)^2 - \tfrac12 \zeta J^2\right)
\end{equation}
are added to the action with $J$ the source used to couple the operator to the theory. The term in $J^2$ is necessary here for renormalizability of the connected-diagram-generating functional $W(J)$ and, subsequently, of the
associated 1PI-diagram generating functional $\Gamma$, aka.~the effective action. Here $\zeta$ is a new coupling constant whose determination we will discuss later. In the physical vacuum, corresponding to $J\to0$, it should decouple again, at least if we were to do the computations exactly. At (any) finite order, $\zeta$ will be explicitly present, even in physical observables, making it necessary to choose it as wisely as possibly. Notice that $\zeta$ is \emph{not} a gauge parameter as it in fact couples to the BRST invariant quantity $J^2$. Indeed, in a BRST invariant theory, we expect the gauge parameter to explicitly cancel order per order from physical observables, a fact guaranteed by \emph{e.g.}~the Nielsen identities \cite{Nielsen:1975fs}, which are in themselves a consequence of BRST invariance \cite{Piguet:1984js}.

Thanks to $\zeta$, the Lagrangian remains now multiplicatively renormalizable (see \citeapp).

To actually compute the effective potential, it is computationally simplest to rely on Jackiw's background field method \cite{Jackiw:1974cv}. Before integrating over any fluctuating quantum fields, a Legendre transform is performed, so that formally $\sigma=\frac12 (a_\mu^h - \bar A_\mu)^2-\zeta J$. Plugging this into the Legendre transformation between $\Gamma$ and $W$, we find that we could just as well have started from the action \eqref{nieuweactie2} with the following unity inserted into the path integral:\footnote{This is equivalent to a Hubbard--Stratonovich transformation, see for instance \cite{Verschelde:2001ia,prep}. This also evades the interpretational issues for the energy when higher-than-linear-terms in the sources are present.}
\begin{equation}\label{uniteit2}
1 = \mathcal N \int[\mathcal D\sigma] \exp-\frac1{2\zeta} \int d^dx \left(\sigma + \tfrac12(a_\mu^h - \bar A_\mu)^2\right)^2 \;,
\end{equation}
with $\mathcal N$ an irrelevant constant. Of course, if we could integrate the path integral exactly, this unity would not change a thing. The situation only
gets interesting if the perturbative dynamics of the theory would prefer to assign a non-vanishing vacuum expectation value to $\sigma$. As such, this $\sigma$ field allows one to include potential non-perturbative information through its vacuum expectation value. In the case without a background, $\sigma$ does indeed condense and a vacuum with $\Braket\sigma\not=0$ is preferred.

For the record, BRST invariance is ensured if we assign {$s\sigma=-s\left(\tfrac12(a_\mu^h - \bar A_\mu)^2\right)$,} which implies off-shell that $s\sigma=0$ thanks to the BRST invariance of $a_\mu^h-\bar A_\mu$.

For completeness, let us write down the full gauge-fixed action here, {(we consider the $\alpha\to 0$ limit right away, that is the Landau-deWitt gauge)}
\begin{eqnarray}\label{nieuweactie3}
S_{\mbox{{\tiny full}}} &=& S_{\mbox{\tiny LDW}}+S_h-\frac1{2\zeta} \left(\sigma + \tfrac12(a_\mu^h - \bar A_\mu)^2\right)^2\\&&\hspace{-1cm}=\int d^4x \left(\frac14(F_{\mu\nu}^a)^2 + ib^a\bar D_\mu(a_\mu^a-\bar A_\mu^a) + \bar c^a\bar D_\mu D_\mu c^a+i \tau^a \bar D_\mu^{ab} (a_\mu^{h,b}-\bar A_{\mu}^b) + \bar\eta^a \bar D_\mu^{ab} D_\mu^{bc}[a^h]\eta^c-\frac1{2\zeta} \left(\sigma + \tfrac12(a_\mu^h - \bar A_\mu)^2\right)^2\right) \nonumber\,.
\end{eqnarray}

The outcome of Sect.~II.D can be immediately generalized, leading to
\begin{equation}\label{stap1bis}
   \Braket{O_1(x_1)\ldots O_n(x_n)}_{S_{\mbox{{\tiny full}}}}=\Braket{O_1(x_1)\ldots O_n(x_n)}_{S_{\mbox{{\tiny mLDW}}}}
\end{equation}
for gauge invariant operators $O_i(x_i)$, where
\begin{equation}\label{stap2bis}
   S_{\mbox{{\tiny mLDW}}}\equiv S_{\mbox{{\tiny LDW}}}-\frac{1}{2\zeta}\int d^dx \left(\sigma + \tfrac12 A_\mu^2\right)^2\;.
\end{equation}
Notice that on-shell, or to be more precise when the $\tau$-~and $b$-equation of motion are used, we have {$s\sigma=-s\left(\tfrac12(a_\mu^h - \bar A_\mu)^2\right)\to -s\left(\tfrac12(a_\mu - \bar A_\mu)^2\right)= -A_\mu^a s A_\mu^a\neq0$,} ensuring that \eqref{stap2bis} is still BRST invariant.

We still need to discuss the new coupling $\zeta$. First note that, given the BRST invariance of the action, we can work in a preferred gauge, that is, the Landau--DeWitt gauge, see Subsection \ref{speciaal}, the conclusions whereof are not effected by the inclusion of the BRST invariant unity  \eqref{uniteit}.

It is evident that $\zeta$ can be interpreted as a genuine new coupling constant. Therefore, we now have two coupling constants, $g^2$ and $\zeta$, with $g^2$ running as usual, that is: independently of $\zeta$. This makes our situation suitable for the Zimmermann reduction of couplings program \cite{Zimmermann:1984sx}, see also \cite{Heinemeyer:2019vbc} for a recent overview. In this program, one coupling ($\zeta$ in our case) is re-expressed as a series in the other (here $g^2$), so that the running of $\zeta$ controlled by $\zeta(g^2)$ is then automatically satisfied, see also \cite{prep}. More specifically, $\zeta(g^2)$ is determined such that the generating functional of connected Green functions, $W(J)$, obeys a standard, linear renormalization group equation \cite{Verschelde:2001ia}.

This selects one consistent coupling $\zeta(g^2)$ from a whole space of allowed couplings, and it is also the unique choice compatible with multiplicative renormalizability \cite{Verschelde:2001ia}. Given that we already pointed out that $\zeta$ should, in principle, not affect physics, we can safely rely here on this special choice, already made earlier in \emph{e.g.} \cite{Verschelde:2001ia}. This choice seems also to be a natural one from the point of view of the loop expansion of the background potential to be used below.\footnote{The $\zeta$-independence of expectation values not involving the $\sigma$ field relies however taking into account the unity in an exact fashion, see \cite{prep}. Clarifying how approximations break this $\zeta$-independence and how increasing the truncation order reduces this dependence remains an interesting question.} In the $\MSbar$ scheme, one finds \cite{Verschelde:2001ia,7}
\begin{subequations} \label{zetadeltazeta}
\begin{gather}
\zeta = \frac{N^2-1}{g^2N} \left(\frac{9}{13} + \frac{g^2N}{16\pi^2}\frac{161}{52} + \mathcal O(g^4)\right)\;, \\
Z_\zeta = 1 - \frac{g^2N}{16\pi^2} \frac{13}{3\epsilon} + \mathcal O(g^2)\;, \\
Z_J = 1 - \frac{Ng^2}{16\pi^2} \frac{35}{6\epsilon} + \mathcal O(g^2)\;,
\end{gather}
\end{subequations}
where $Z_\zeta$, $Z_J$ are the renormalization factors of $\zeta J^2$, $J$ respectively.

\subsection{Introduction of asymmetry in the $d=2$ gluon condensate} \label{asym}
When temperature is switched on, it is natural to consider the timelike and spacelike components of the $A^h A^h$ condensate separately, or equivalently to introduce the BRST invariant electric-magnetic asymmetry \cite{Chernodub:2008kf}:
\begin{equation}
	\Delta_{A^2} = \langle g^2A_0^h A_0^h \rangle - \frac13 \langle g^2A_i^h A_i^h \rangle \;,
\end{equation}
where the Latin index denotes the space components and $A_\mu^h$ is a shorthand for $a_\mu^h-\bar A_\mu$. This asymmetry can be included in exactly the same way as the $(a_\mu^h-\bar A_\mu)^2$ condensate, namely we add
\begin{equation} \label{lcobg}
\int d^dx \left(\tfrac12 K_{\mu\nu}  ((a_\mu^h - \bar A_\mu)(a_\nu^h - \bar A_\nu) - \tfrac{\delta_{\mu\nu}}{d} (a_\mu^h - \bar A_\mu)^2 ) - \tfrac12 \omega K_{\mu\nu}K_{\mu\nu}+\frac{\omega}{2d}K_{\mu\mu}^2\right)\;,
\end{equation}
where the dimension-two symmetric source $K_{\mu\nu}$ couples to the traceless operator $((a_\mu^h - \bar A_\mu)(a_\nu^h - \bar A_\nu) - \tfrac{\delta_{\mu\nu}}{d} (a_\mu^h - \bar A_\mu)^2 )$, see also \cite{Dudal:2009tq}.

The same goal can be reached by directly adding an extra part to the action:
\begin{equation}
	\frac1{2\omega} \int d^dx \left( \varphi_{\mu\nu} + \frac12 (a_\mu^h - \bar A_\mu)(a_\nu^h - \bar A_\nu) \right)^2 \;,
\end{equation}
with $\varphi_{\mu\nu}$ an auxiliary field analogous to $\sigma$, but which we will take to be a traceless matrix and which will thus couple to the asymmetric part of the condensate. The parameter $\omega$ is the analogon of $\zeta$. As we are interested in the asymmetry, we parameterize the mass matrix as
\begin{equation} \label{param}
	\varphi_{\mu\nu} = \omega \mathbb{A} \begin{pmatrix} 1 &&& \\ &-\frac1{d-1}&& \\&&\ddots&\\ &&&-\frac1{d-1} \end{pmatrix} \;,
\end{equation}
\emph{i.e.} we preserve rotational invariance in the spatial part. Determining $\omega$ in the same way we found $\zeta$ gives \cite{Dudal:2009tq}
\begin{subequations} \label{omegadeltaomega}
\begin{gather}
	\omega = \frac{N^2-1}{g^2N} \left(\frac{1}{4} + \frac{73}{1044} \frac{g^2N}{16\pi^2}  + \mathcal O(g^4)\right)\;,\\
Z_\omega = 1 + \frac{g^2N}{24\pi^2} \frac{11}{\epsilon} + \mathcal O(g^2)\;, \\
Z_K = 1 - \frac{g^2N}{16\pi^2} \frac{29}{6\epsilon} + \mathcal O(g^2)\;,
\end{gather}
\end{subequations}
to one-loop order. Here, $Z_\omega$, $Z_K$ are the renormalization factors of $K_{\mu\nu}K_{\mu\nu}-\tfrac{1}{2d}K_{\mu\mu}^2$, $K_{\mu\nu}$ respectively.

\subsection{Background field independence of physical observables}
Using the standard Landau--DeWitt gauge condition, it can be nicely shown using the (extended) Slavnov--Taylor identity that physical observables do not depend on the choice of the background $\bar A(x)$, which is of course expected given that choosing $\bar A(x)$ corresponds to choosing a specific gauge. For a formal proof, see \cite{Ferrari:2000yp}. The crux of the matter is to extend the BRST operator $s$ to also act on the background field via $s\bar A=\Omega$, $s\Omega=0$ (see also \citeapp, based on \cite{brstbackground}), with $\Omega$ an auxiliary Grassmann background field that is to be sent to zero again eventually. This is actually an extension of the BRST method to formally prove gauge parameter independence of observables in linear gauges, see \cite{Piguet:1984js,Piguet:1995er}, in which case the gauge parameter $\alpha$ is also made part of a BRST doublet.

We will not exactly follow this procedure here, see also the comment below \eqref{nieuweactie2}. Indeed, we would also need to properly extend the BRST operator to auxiliary fields to maintain a full extended BRST invariance of the action. But we can follow a slightly different route to make our point, again benefitting from the ``reduction'' $a^h\to a$ in the Landau--DeWitt gauge.

Let us first consider observables that are not directly depending on $a$, such as the partition function, free energy and related quantities.  One finds, only writing the integration over the gluon degrees of freedom for simplicity,
\begin{eqnarray}
\frac{\delta}{\delta\bar A(x)}\int \left[{\cal D}a\right] e^{-S_{\mbox{{\tiny LDW}}} -\frac{1}{2\zeta}\int d^dx \left(\sigma +\frac{1}{2}(a-\bar A)^2\right)^2} & = & \frac{\delta}{\delta\bar A(x)}\int \left[{\cal D}A\right]e^{-S_{\mbox{{\tiny LDW}}} -\frac{1}{2\zeta}\int d^dx \left(\sigma +\frac{1}{2}A^2\right)^2} \\&=&\int \left[{\cal D}A\right]s\left(\frac{\delta}{\delta\bar A(x)}\left(\int d^d z\bar c \bar D A\right)\right)e^{-S_{\mbox{{\tiny LDW}}} -\frac{1}{2\zeta}\int d^dx \left(\sigma +\frac{1}{2}A^2\right)^2}=0 \nonumber
\end{eqnarray}
using the BRST symmetry of the vacuum/action, including of the last term of the action (the unity), see the comments below \eqref{stap2bis}, {and the fact that the gauge-fixing part of $S_{\mbox{\tiny LDW}}$ is BRST-exact.} We also changed integration variable $a\to A$.

Using a slightly different argumentation, we can extend the argument to correlation functions of (renormalizable) gauge invariant operators $O_i(x_i)$. We assume $x_i\neq x_j$ for $i\neq j$, as otherwise we are looking at a different composite quantum operator that needs it separate renormalization. We get from \eqref{stap1bis}
\begin{eqnarray}\label{u1}
  \frac{\delta}{\delta \bar A(x)}\Braket{O_1\ldots O_n}_{S_{\mbox{{\tiny full}}}}&=&\frac{\delta}{\delta \bar A(x)}\int [{\mathcal{D}a\mathcal{D}\sigma}] O_1\ldots O_n e^{-S_{\mbox{{\tiny LDW}}} -\frac{1}{2\zeta}\int d^dx \left(\sigma +\frac{1}{2}(a-\bar A)^2\right)^2}\nonumber\\
  &&\hspace{-4cm}= \Braket{O_1\ldots O_n\frac{\delta}{\delta\bar A(x)}\left(s\int d^d z\bar c \bar D A\right) }_{S_{\mbox{{\tiny mLDW}}}}+  \int [{\mathcal{D}a\mathcal{D}\sigma}] O_1\ldots O_n \frac{1}{2\zeta}(a-\bar A)\left(\sigma+\frac{1}{2}(a-\bar A)^2\right)e^{-S_{\mbox{{\tiny LDW}}} -\frac{1}{2\zeta}\int d^dx \left(\sigma +\frac{1}{2}(a-\bar A)^2\right)^2}\nonumber\\
  &&\hspace{-4cm}= \Braket{s\left(O_1\ldots O_n\frac{\delta}{\delta\bar A(x)}\left(\int d^d z\bar c \bar D A\right) \right)}_{S_{\mbox{{\tiny mLDW}}}}- \int [{\mathcal{D}a\mathcal{D}\sigma}] O_1\ldots O_n (a-\bar A)\frac{\delta}{\delta \sigma}e^{-S_{\mbox{{\tiny LDW}}} -\frac{1}{2\zeta}\int d^dx \left(\sigma +\frac{1}{2}(a-\bar A)^2\right)^2}\nonumber\\
  &=&\int [{\mathcal{D}a\mathcal{D}\sigma}] \frac{\delta}{\delta \sigma}\left(O_1\ldots O_n \right) (a-\bar A)e^{-S_{\mbox{{\tiny LDW}}} -\frac{1}{2\zeta}\int d^dx \left(\sigma +\frac{1}{2}(a-\bar A)^2\right)^2}=0\;,
  \end{eqnarray}
where the one-but-last step is again based on BRST invariance (first term) {and the functional version of the trivial identity $(x+y)e^{-(x+y)^2/2}=-de^{-(x+y)^2/2}/dx$ (second term).  In the last step, one recognizes the (functional) integral of a (functional) total derivative, which vanishes in the absence of boundary terms. We mention that the present argumentation does not require using the unity (\ref{uniteit2}) and, therefore, shows that the background-independence property should be relatively robust to practical expansions of (\ref{uniteit2}) as used below.}

\subsection{Background gauge invariance} \label{bgi}
Besides BRST invariance, another important symmetry when working with a gluonic background field is background gauge invariance:
\begin{equation}
	\delta \bar A_\mu^a = \bar D_\mu^{ab} \beta^b \;, \qquad \delta\varphi^a = -f^{abc} \beta^b \varphi^c \;,
\end{equation}
where $\varphi^a$ stands for all the quantum fields. The ordinary Yang--Mills action with the Faddeev--Popov ghost part is invariant under this symmetry. To see the invariance of the extra part \eqref{lcobg}, consider the expansion \eqref{ahexp}. The background on the left-hand side of that expression only appears in covariant derivatives, such that the entire expression transforms as a matter field:
\begin{equation}\label{achter}
	\delta((a^h)_\mu^a - \bar A_\mu^a) = -f^{abc} \beta^b ((a^h)_\mu^c - \bar A_\mu^c) \;,
\end{equation}
meaning the mass term \eqref{lcobg} is invariant.

In order to use the background field formalism, we can now follow the arguments of \cite{Reinosa:2014ooa}. Consider the quantum effective action of the gluon field computed in the presence of a background $\bar A_\mu^a$: $\Gamma_{\bar A}[a]$. The physical vacuum is found by minimizing with respect to $a_\mu^a$:
\begin{equation} \label{minim}
	\Gamma_{\bar A}[a^\text{cl}_{\bar A}] \leq \Gamma_{\bar A}[a] \qquad \forall a_\mu^a \;,
\end{equation}
where $(a^\text{cl}_{\bar A})_\mu^a = \Braket{a_\mu^a}_{\bar A}$ is the value of $a_\mu^a$ in this minimum. Now, thanks to BRST invariance, the background is in essence a gauge parameter, such that physical quantities may not depend on it, and we can freely choose it\footnote{This property is fragile to the use of approximations or modelling (as those considered within nonperturbative approaches or the Curci--Ferrari model) leading to potential spurious effects in the results obtained using the background effective action. Recently, an alternative approach has been put forward that relies, instead, on using the standard effective action $\Gamma_{\bar A_c}[A]$ in a particular gauge, as defined by the choice of a center-symmetric background $\bar A_c$. The rationale for using this approach does not rely on the background independence of the free-energy and is thus more robust to violations of the latter. In the present BRST-invariant loop expanded approach, we expect these violations to be minimal and the two approaches to be essentially equivalent. This will be investigated elsewhere.}. Thence choosing a self-consistent background $\bar A_s$ defined by the condition $\bar A_s=a^\text{cl}_{\bar A_s}$, we find
\begin{equation}\label{uu}
	\Gamma_{\bar A_s}[\bar A_s]=\Gamma_{\bar A_s}[a^\text{cl}_{\bar A_s}] = \Gamma_{\bar A}[a^\text{cl}_{\bar A}] \leq \Gamma_{\bar A}[\bar A] \;,
\end{equation}
where the first equality follows from the self-consistency condition, the second one is BRST invariance and the third one is the minimization condition \eqref{minim} for the specific value $a_\mu^a = \bar A_\mu^a$. In conclusion, the minimum of the quantum effective action can be found by minimizing $\Gamma_{\bar A}[\bar A]$ with respect to $\bar A_\mu^a$. Thanks to the remaining background gauge invariance, we also know that $\Gamma_{\bar A}[\bar A]$ will be a (background) gauge invariant functional of $\bar A$. In the presence of the condensate and the asymmetry, the background effective action is also a function of the condensate and the asymmetry, variables with respect to which it also needs to be minimized.

It is important to note that $\Gamma_{\bar A}[\bar A]$ does not need to be $\bar A$ independent and its explicit computation in the next section will make this dependence quite explicitly clear.\footnote{{It can be shown, however, to be constant on constant backgrounds, at zero temperature \cite{Reinosa:2015gxn,Reinosa:2020mnx}.}} This is not at odds with the previous subsection. On the contrary, it is even to be expected, as $\Gamma_{\bar A}[\bar A]$ does not obey a Slavnov--Taylor identity to begin with. Indeed, to avoid misconceptions, we stress here that the functional $\Gamma_{\bar A}[\bar A]$ is \emph{not} the standard quantum effective action generating 1PI graphs, which is $\Gamma_{\bar A}[\bar A]$. It is however still a useful functional that also appears in the so-called ``background equivalence theorem'' for which we refer to \emph{e.g.}~\cite{Ferrari:2000yp} for more details and references. Here, we appreciate its usefulness to select self-consistent backgrounds from its minimization, leading to estimates of the ground state free energy, based on \eqref{uu}.

\section{Computation of the background effective potential}\label{sec:pot}

In what follows, we evaluate the background field effective potential whose minima give access to the self-consistent background and thus to order parameters for the confinement/deconfinement transition. We work in the Landau--DeWitt gauge which means that we send $\alpha$ to $0$. As we have explained, we can then consider replacing $a^h$ by $a$ which considerably simplifies the calculations.

\subsection{Warming up in the absence of asymmetry}
For the sake of simplicity, let us assume first that there is no asymmetry. This approximation will turn out to be justified as the asymmetry we will find below is tiny.
In order to evaluate the background effective potential at one-loop order, we need the terms in the action that are at most quadratic in the fields:
\begin{eqnarray}
	&&\int d^dx \Bigg(\frac\zeta2 m^4 \left(1-\frac{\delta\zeta}\zeta\right) + \frac{1}{2}A_\mu^a \left(- \delta_{\mu\nu} \bar D^2_{ab} + \left(1-\frac1\alpha\right) \bar D_\nu^{ac} \bar D_\mu^{cb} + \delta^{ab}\delta_{\mu\nu} m^2 \right) A_\nu^b+\bar c^a \bar D^2_{ab} c^b \Bigg),
\end{eqnarray}
where the limit $\alpha\to0$ is assumed, and where we used the notation $Z_\zeta = 1-\delta\zeta/\zeta$. Renormalization factors in the part quadratic in the quantum fields are ignored, as they will not be necessary at one-loop order. We also wrote $\sigma=\zeta m^2$.

Integrating out the fields yields traces of logarithms of the operators multiplying their quadratic parts. In order to deal with them, we work in a space where the covariant derivative is diagonal. Let us see how this works in the SU(2) case before generalizing to $SU(N)$. We first go over to a basis in color space where $\epsilon^{ab3}$ is diagonal:
\begin{equation}
	\epsilon^{ab3} \mx e_\kappa^b = i\kappa \mx e_\kappa^a \;,
\end{equation}
with $\kappa\in\{-1,0,+1\}$. If we write $A^a_\mu = A_\mu^\kappa \mx e^\kappa_a$, then we immediately find that $\bar D_0 A_\mu^\kappa = \partial_0 A_\mu^\kappa - ir\kappa T A_\mu^\kappa$ (no sum over $\kappa$). In Fourier space we can therefore write
\begin{equation}
	\bar D_\mu A_\nu^\kappa = i P_\mu^\kappa A_\nu^\kappa \qquad  (P_0^\kappa = p_0 - r\kappa T \;, \quad P_i^\kappa = p_i) \;.
\end{equation}
As such the one-loop effective potential is
\begin{equation}
	V(r, m^2)=\frac\zeta2 m^4 \left(1-\frac{\delta\zeta}\zeta\right)+ \frac12 \tr\ln \left( \delta_{\mu\nu} P^2_\kappa - \left(1-\frac1\alpha\right) P_\mu^\kappa P_\nu^\kappa + \delta_{\mu\nu} m^2 \right) - \tr\ln P^2_\kappa \;,
\end{equation}
where the trace refers to space-time indices, as well as color charges and momenta.
An operator of the type
\begin{equation}
	X\delta_{\mu\nu}+Y\frac{P^\kappa_\mu P^\kappa_\nu}{P^2_\kappa}
\end{equation}
has one eigenvector parallel to $P^\kappa_\mu$ with eigenvalue $X+Y$, and $d-1$ eigenvectors perpendicular to $P^\kappa_\mu$ with eigenvalue $X$. This means that
\begin{equation}
	\tr\ln \left(X\delta_{\mu\nu}+Y\frac{P^\kappa_\mu P^\kappa_\nu}{P^2_\kappa}\right) = (d-1)\ln X+\ln (X+Y) \;.
\end{equation}
Using this, we arrive at
\begin{equation}
V(r,m^2)=\frac\zeta2 m^4 \left(1-\frac{\delta\zeta}\zeta\right)+\frac{d-1}2 \tr\ln (P^2_\kappa + m^2)+\frac12 \tr\ln \left(\frac{P^2_\kappa}{\alpha} + m^2\right)-\tr\ln P^2_\kappa\,,
\end{equation}
where the trace now refers to the color charges and the momenta. In the limit $\alpha\to0$ neglecting a trivial term, this gives
\begin{equation}
V(r,m^2)=\frac\zeta2 m^4 \left(1-\frac{\delta\zeta}\zeta\right)+\frac{d-1}2 \tr\ln (P^2_\kappa + m^2)-\frac{1}{2} \tr\ln P^2_\kappa\,,\label{eq:su2_pot}
\end{equation}
where the last term comes from a partial cancellation between the ghost contribution and the massless longitudinal gluon mode. The analysis of the SU(2) potential can be restricted to the interval $r\in [0,2\pi]$ and even $[0,\pi]$, the center-symmetric point corresponding to $r=\pi$.

The formula for the background potential in the $SU(N)$ case is formally the same as (\ref{eq:su2_pot}). The only change is that the $N^2-1$ labels $\kappa$ become vectors of $\mathds{R}^{N-1}$ whose components are denoted $\kappa_j$, with $j$ referring to the diagonal directions of the algebra ($3$ and $8$ in the SU(3) case for instance). Correspondingly, the variable $r$ also becomes a vector of $\mathds{R}^{N-1}$ of components $r_j$, and we have now $P_0^\kappa=p_0 - r_j\kappa_j T$. Out of the $N^2-1$ labels $\kappa$,  $N-1$ are equal to $0$ and the rest are the roots characterizing the associated Lie algebra. In the case of SU(3) for instance, there are two zeros and six roots $\pm(1,0)$, $\pm (1/2,\sqrt{3}/2)$ and $\pm (1/2,-\sqrt{3}/2)$. The analysis of the SU(3) potential can be restricted to $r_3\in [0,2\pi]$ while charge conjugation invariance (in pure YM) imposes $r_8=0$ (in this range of values for $r_3$). The center-symmetric point corresponds in this case to $r_3=4\pi/3$. More on this can be found in \emph{e.g.}~\cite{Reinosa:2020mnx}. We shall later exploit these remarks to infer the expression for the SU(3) potential from that of the SU(2) potential.

\subsection{Including the asymmetry}
In the presence of the asymmetry, the quadratic part of the action reads
\begin{eqnarray}
	&&\int d^dx \Bigg(\frac\zeta2 m^4 \left(1-\frac{\delta\zeta}\zeta\right) + \frac\omega2 \mathbb{A}^2 \frac d{d-1} \left(1-\frac{\delta\omega}\omega\right) + {\frac{1}{2}}A_\mu^a \left(- \delta_{\mu\nu} \bar D^2_{ab} + \left(1-\frac1\alpha\right) {\bar D_\nu^{ac} \bar D_\mu^{cb}} + \delta^{ab}\delta_{\mu\nu} m^2 + \delta^{ab}M_{\mu\nu} \right) A_\nu^b\nonumber\\ &&+ \bar c^a \bar D^2_{ab} c^b \Bigg) \;,
\end{eqnarray}
where, again, the limit $\alpha\to0$ is assumed, and where we used the notation $Z_\omega$ analogous to $Z_\zeta = 1-\frac{\delta\zeta}\zeta$. We also wrote $\varphi_{\mu\nu} = \omega M_{\mu\nu}$.

Integrating out the fields, we arrive now at the SU(N) one-loop effective potential
\begin{eqnarray}
V(r,m^2,A) & = & \frac\zeta2 m^4 \left(1-\frac{\delta\zeta}\zeta\right) + \frac\omega2 \mathbb{A}^2 \frac d{d-1} \left(1-\frac{\delta\omega}\omega\right)\nonumber\\
& + &  \frac12 \tr\ln \left( \delta_{\mu\nu} P^2_\kappa - \left(1-\frac1\alpha\right) P_\mu^\kappa P_\nu^\kappa + \delta_{\mu\nu} m^2 + M_{\mu\nu} \right) - \tr\ln P^2_\kappa \;,
\end{eqnarray}
where, as before, a summation over $\kappa$ is implied. In order to more easily handle the $d\times d$ matrix coming from the gluon fields, let us, following \cite{Dudal:2009tq}, separate out the part without the mass matrix $M_{\mu\nu}$:
\begin{multline}
	\tr\ln \left( \delta_{\mu\nu} P^2_\kappa - \left(1-\frac1\alpha\right) P_\mu^\kappa P_\nu^\kappa + \delta_{\mu\nu} m^2 + M_{\mu\nu} \right) \\
	= \tr\ln \left( \delta_{\mu\nu} P^2_\kappa - \left(1-\frac1\alpha\right) P_\mu^\kappa P_\nu^\kappa + \delta_{\mu\nu} m^2 \right) + \tr\ln \left( \delta_{\mu\nu} + \frac1{P^2_\kappa + m^2} \left(\delta_{\mu\lambda} - (1-\alpha) \frac{P_\mu^\kappa P_\lambda^\kappa}{P^2_\kappa+\alpha m^2}\right) M_{\lambda\nu} \right) \;.
\end{multline}
In the limit $\alpha\to0$ this gives
\begin{equation}
	(d-1) \tr\ln (P^2_\kappa + m^2) + \tr\ln P^2_\kappa + \tr\ln \left( \delta_{\mu\nu} + \frac1{P^2_\kappa + m^2} \left(\delta_{\mu\lambda} - \frac{P_\mu^\kappa P_\lambda^\kappa}{P^2_\kappa}\right) M_{\lambda\nu} \right)
\end{equation}
plus an irrelevant constant term.

If we now consider the operator in the last term, we can write it in an orthonormal basis consisting of the vectors
\begin{itemize}
	\item the unit vector pointing in the direction of $P_\mu^\kappa$,
	\item the vector obtained by replacing the timelike component of $P_\mu^\kappa$ by $-p_i^2/P_0^\kappa$(so as to make a vector perpendicular to $P_\mu^\kappa$) followed by norming this vector,
	\item an orthonormal basis of spacelike vectors perpendicular to $p_i$.
\end{itemize}
In this basis, the operator under consideration is lower triangular, such that its determinant is the product of its diagonal elements. These diagonal elements are found to be
\begin{itemize}
	\item 1
	\item $1+\frac {\mathbb{A}}{P_\kappa^2+m^2} \left(1-\frac d{d-1} \frac{(P_0^\kappa)^2}{P_\kappa^2}\right)$,
	\item $1-\frac {\mathbb{A}}{P_\kappa^2+m^2} \frac 1{d-1}$ (with multiplicity $d-2$).
\end{itemize}

Gathering all these results, we find that the effective potential at one loop is equal to
\begin{eqnarray}
V(r,m^2,A) & = & \frac\zeta2 m^4 \left(1-\frac{\delta\zeta}\zeta\right) + \frac\omega2 \mathbb{A}^2 \frac d{d-1} \left(1-\frac{\delta\omega}\omega\right) - \tr\ln P^2_\kappa\nonumber\\
 & + & \frac12\tr\ln\left( P_\kappa^2(P_\kappa^2 + m^2) + \mathbb{A} \left(P_\kappa^2-\frac d{d-1} (P_0^\kappa)^2\right) \right) + \frac{d-2}2 \tr\ln\left( P_\kappa^2 + m^2 - \frac {\mathbb{A}}{d-1} \right).\label{eq:pot_A}
\end{eqnarray}
As a cross-check of this formula, in~\citeappvier~we provide an alternative derivation within the Nakanishi-Lautrup formalism. We also notice that, upon taking the limit $\smash{A\to 0}$ one retrieves Eq.~(\ref{eq:su2_pot}).

\subsection{Evaluation of the (sum-)integrals}
The formula (\ref{eq:su2_pot}) and its generalization (\ref{eq:pot_A}) involve various (sum-integrals) which we now evaluate. We consider first the SU(2) case for simplicity, and then use it to infer the corresponding SU(3) formulas.

The expression in the last logarithm of (\ref{eq:pot_A}) expands as $(p_0-r\kappa T)^2+\vec p^2 + m^2-\frac {\mathbb{A}}{d-1}$, and we can immediately apply formula \eqref{finitetformula} from~\citeappdrie~ to find that
\begin{multline}
	\frac{d-2}2 \tr\ln\left( P_\kappa^2 + m^2 - \frac{\mathbb{A}}{d-1} \right) = 3 \frac{d-2}2 \tr_{T=0} \ln\left( p^2 + m^2 - \frac{ \mathbb{A}}{d-1} \right) \\ + 2T \int \frac{d^3p}{(2\pi)^3} \ln\left(1 - 2e^{-\frac{\sqrt{\vec p^2+m^2-\frac{\mathbb{A}}3}}T} \cos r + e^{-2\frac{\sqrt{\vec p^2+m^2-\frac{\mathbb{A}}3}}T}\right) + T \int \frac{d^3p}{(2\pi)^3} \ln\left(1 - e^{-\frac{\sqrt{\vec p^2+m^2-\frac {\mathbb{A}}3}}T}\right)^2 \;.
\end{multline}
Here we put $d=4$ in the (finite) zero-temperature correction part, and we summed over $\kappa$. In the limit $\mathbb{A}\to0$, this formula can be adapted to obtain the integral needed in (\ref{eq:su2_pot}). Moreover, when both $\mathbb{A}=0$ and $m=0$, it can be adapted to obtain
\begin{equation}
	- \tr\ln P^2_\kappa = - 3 \tr_{T=0}\ln p^2 - 2T \int \frac{d^3p}{(2\pi)^3} \ln\left(1 - 2e^{-\frac{\sqrt{\vec p^2}}T} \cos r + e^{-2\frac{\sqrt{\vec p^2}}T}\right) - T \int \frac{d^3p}{(2\pi)^3} \ln\left(1 - e^{-\frac{\sqrt{\vec p^2}}T}\right)^2 \;,
\end{equation}
which also appears in (\ref{eq:pot_A}). Finally, when it comes to the expression in the one-to-last logarithm of (\ref{eq:pot_A}), it writes $(p_0-r\kappa T)^4 +(p_0-r\kappa T)^2 (2\vec p^2+m^2-\frac{ \mathbb{A}}{d-1}) + \vec p^2 (\vec p^2+m^2+\mathbb{A})$. With the definitions
\begin{subequations} \begin{gather}
	\alpha = 2\vec p^2 + m^2 - \frac{\mathbb{A}}{d-1} \;, \qquad \beta = \vec p^2(\vec p^2+m^2+\mathbb{A}) \;, \\
	z_\pm = \frac{\alpha\pm\sqrt{\alpha^2-4\beta}}2 \;,
\end{gather} \end{subequations}
this becomes $((p_0-r\kappa T)^2+z_+) ((p_0-r\kappa T)^2+z_-)$. As such we find
\begin{multline}
	\frac12\tr\ln\left( P_\kappa^2 + m^2 + \mathbb{A} \left(1-\frac d{d-1} \frac{(P_0^\kappa)^2}{P_\kappa^2}\right) \right) + \frac12 \tr\ln P^2_\kappa \\
	= \frac32\tr_{T=0} \ln\left( p^2 + m^2 + \mathbb{A} \left(1-\frac d{d-1} \frac{p_0^2}{p^2}\right) \right) + \frac32 \tr_{T=0}\ln p^2 \\
	+ T \int \frac{d^3p}{(2\pi)^3} \ln\left(1 - 2e^{-\frac{\sqrt{z_+}}T} \cos r + e^{-2\frac{\sqrt{z_+}}T}\right) + \frac T2 \int \frac{d^3p}{(2\pi)^3} \ln\left(1 - e^{-\frac{\sqrt{z_+}}T}\right)^2 \\
	+ T \int \frac{d^3p}{(2\pi)^3} \ln\left(1 - 2e^{-\frac{\sqrt{z_-}}T} \cos r + e^{-2\frac{\sqrt{z_-}}T}\right) + \frac T2 \int \frac{d^3p}{(2\pi)^3} \ln\left(1 - e^{-\frac{\sqrt{z_-}}T}\right)^2 \;.
\end{multline}
Putting all of this together, we find that the one-loop effective potential at finite temperature is
\begin{multline}
V(r,m^2,\mathbb{A})=V_{T=0} + T \int \frac{d^3p}{(2\pi)^3} \ln\left(1 - 2e^{-\frac{\sqrt{z_+}}T} \cos r + e^{-2\frac{\sqrt{z_+}}T}\right) + \frac T2 \int \frac{d^3p}{(2\pi)^3} \ln\left(1 - e^{-\frac{\sqrt{z_+}}T}\right)^2 \\
	+ T \int \frac{d^3p}{(2\pi)^3} \ln\left(1 - 2e^{-\frac{\sqrt{z_-}}T} \cos r + e^{-2\frac{\sqrt{z_-}}T}\right) + \frac T2 \int \frac{d^3p}{(2\pi)^3} \ln\left(1 - e^{-\frac{\sqrt{z_-}}T}\right)^2 \\
	+ 2T \int \frac{d^3p}{(2\pi)^3} \ln\left(1 - 2e^{-\frac{\sqrt{\vec p^2+m^2-\frac{\mathbb{A}}3}}T} \cos r + e^{-2\frac{\sqrt{\vec p^2+m^2-\frac{\mathbb{A}}3}}T}\right) + T \int \frac{d^3p}{(2\pi)^3} \ln\left(1 - e^{-\frac{\sqrt{\vec p^2+m^2-\frac{\mathbb{A}}3}}T}\right)^2 \\
	- 2T \int \frac{d^3p}{(2\pi)^3} \ln\left(1 - 2e^{-\frac{\sqrt{\vec p^2}}T} \cos r + e^{-2\frac{\sqrt{\vec p^2}}T}\right) - T \int \frac{d^3p}{(2\pi)^3} \ln\left(1 - e^{-\frac{\sqrt{\vec p^2}}T}\right)^2 \;.\label{eq:final}
\end{multline}
The zero-temperature contribution does not depend on $r$ and is therefore the same as what was found in \cite{Dudal:2009tq}, namely
\begin{eqnarray}\label{effpot}
V_{T=0}  &=& \frac{N^2 - 1}{2 (4\pi)^2} \Biggl\{ \frac{1}{18}  \ln\left( \frac{m^2- \mathbb{A}/3}{\overline{\mu}^2}\right) \left[ 7\mathbb{A}^2 + 27 m^4  \right] +  \left[ -\frac{155}{522} \mathbb{A}^2 + \frac{11}{12} \mathbb{A} m^2 - \frac{87}{26} m^4 + \frac{1}{4} \frac{m^6}{\mathbb{A}} \right] \nonumber\\
&& +\frac{1}{18}\left[5 \mathbb{A}^2 + 12 \mathbb{A} m^2 + 9m^4 \right] \left[ \ln\left( \frac{\mathbb{A}}{\mathbb{A}-3m^2}\right) +\ln\left(1+\sqrt{ \frac{m^2+\mathbb{A}}{m^2-\frac{\mathbb{A}}{3}}}\right)-\ln\left(1-\sqrt{ \frac{m^2+\mathbb{A}}{m^2-\frac{\mathbb{A}}{3}}}\right) \right] \nonumber\\
  &&- \frac{\left(m^2 - \frac{\mathbb{A}}{3}\right)}{12 \mathbb{A}} (6 \mathbb{A}^2 + 11 \mathbb{A} m^2 + 3 m^4) \sqrt{ \frac{m^2+\mathbb{A}}{m^2-\frac{\mathbb{A}}{3}}}  + \frac{9}{13} \frac{(4\pi)^2}{g^2 N} m^4 + \frac{1}{3} \frac{(4\pi)^2}{g^2 N} \mathbb{A}^2  \Biggr\}\;,
\end{eqnarray}
with $N=2$.

The extension to SU(3) is straightforward. Given that we can restrict to $r_8=0$ and that the label $\kappa$ spans now $2$ zeros rather than $1$ together with the SU(3) roots given above, with respective projections along the direction $3$ being $\pm 1$ and twice $\pm 1/2$, the SU(3) formula for the potential as a function of $r\equiv r_3$ can be obtained from (\ref{eq:final}) by: 1) using $V_{T=0}$ with $N=3$; 2) duplicating the integrals that do not depend on $r$; 3) keeping the integrals that depend on $r$; 4) adding twice these integrals with $r\to r/2$.

\section{Numerical results and discussion}\label{sec:results}

\subsection{Zero-temperature limit and parameter setting}
Using Eq.~(\ref{effpot}), it can be checked that one has {$\mathbb{A}=0$} at $T=0$, see Ref.~\cite{Dudal:2009tq}.  Then, solving $\partial V_{T=0}/\partial m^2|_{m^2=m^2_{\min,T=0}}=0$ with the renormalization group optimized choice $\bar\mu=m_{\min,T=0}$, one finds
\begin{equation}
 \frac{g^2(m^2_{\min,T=0})N}{16\pi^2} = \frac{36}{187} \;.
\end{equation}
Using the one-loop $\beta$-function and $\lms\approx0.752\sqrt{\sigma}$ for $N=2$ \cite{Lucini:2008vi}, that is $\lms^{N=2}\approx0.331$~GeV using $\sqrt{\sigma}=0.44$~GeV for the scale setting, one finds as solution
\begin{equation}\label{massa}
m^2_{\min,T=0}= e^{17/12}\lms^{N=2} \approx 0.451~\text{GeV}^2.
\end{equation}
In the SU(3) case, one finds instead
\begin{equation}\label{massa}
m^2_{\min,T=0}=e^{17/12}\lms^{N=3}\approx 0.207~\text{GeV}^2
\end{equation}
based on $\lms^{N=3}\approx 0.224$~GeV \cite{Boucaud:2008gn}.

Interestingly enough, these correspond to values of the mass parameter equal to $672$ MeV and $455$ MeV respectively, pretty close to those obtained when fitting Landau gauge propagators using the Curci--Ferrari model; \cite{Reinosa:2014ooa} used $710$~MeV and $510$~MeV respectively. We will see below that the similarities with the CF model do not end here.

With the parameters fixed at $T=0$, we can now study finite temperature effects and their impact on both the Polyakov loop and the asymmetry.

\begin{figure}[t]
	\centering
	\includegraphics[width=0.4\linewidth]{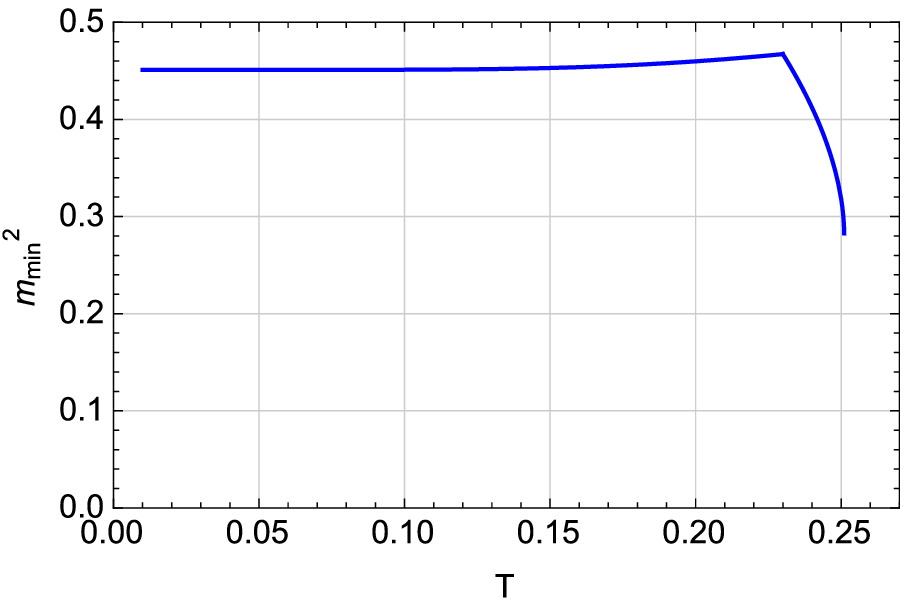}\quad\includegraphics[width=0.4\linewidth]{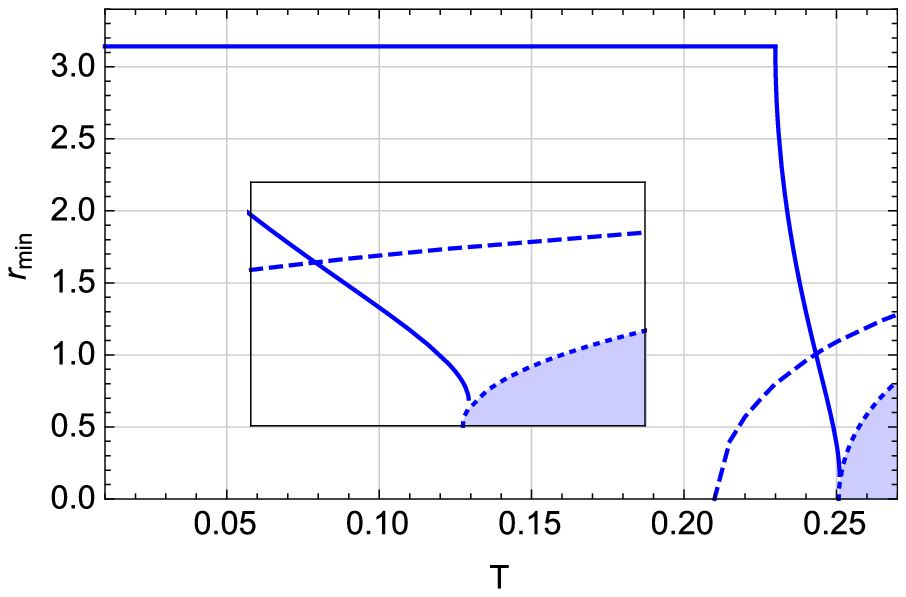}
	\includegraphics[width=0.4\linewidth]{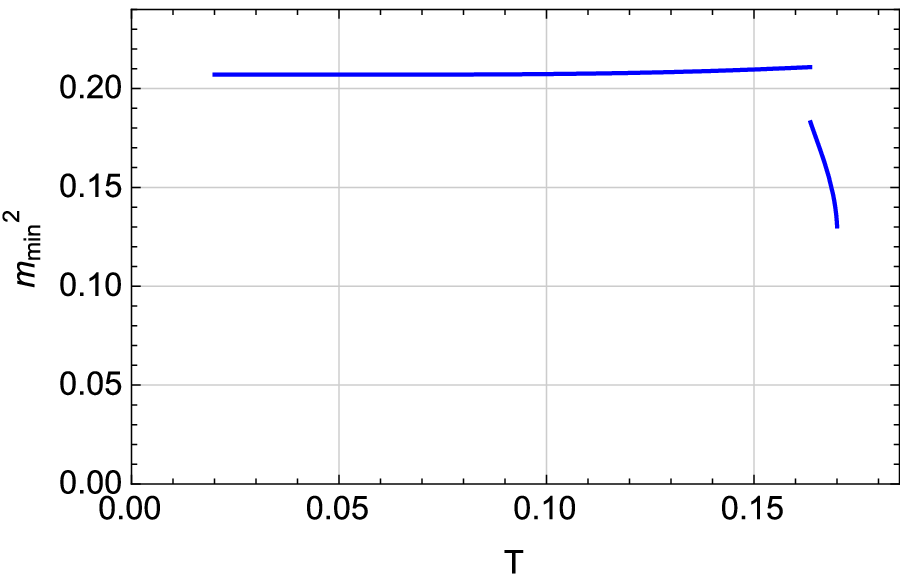}\quad\includegraphics[width=0.4\linewidth]{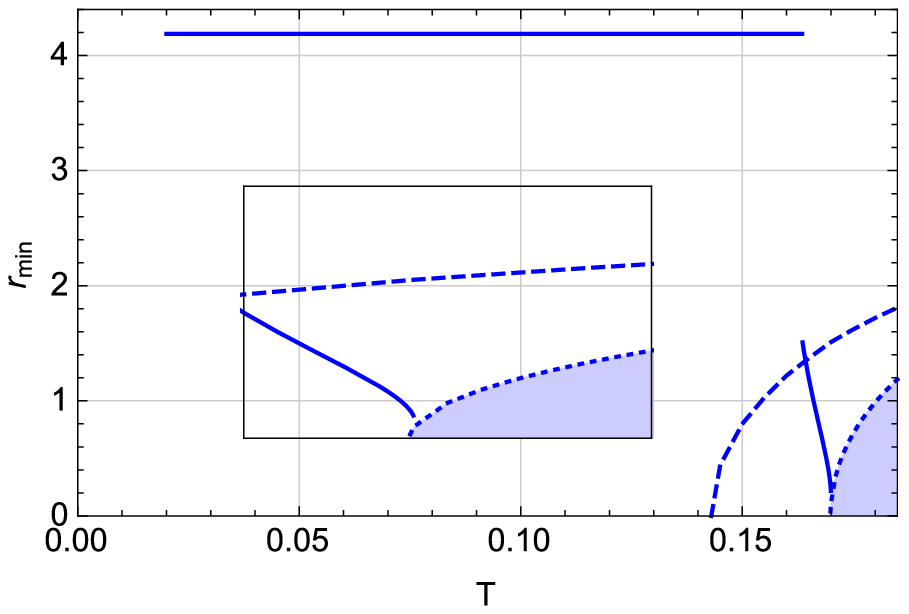}
	\caption{{ The condensate $m^2_{\min}$ (left) and background $r_{\min}$ (right) at the minimum of the potential for the SU(2) (top) and SU(3) (bottom) theories, under the assumption that $\mathbb{A}=0$. The dashed lines in the right plots show the region below which the absolute minimum of the potential is no longer a stationary point although there is still a local stationary minimum. The dotted lines show the region below which the stationary minimum is lost. The inset zooms in on the region above the transition where the solution is lost, see the text below for more details. In units GeV. }}\label{fig:1}
\end{figure}

\subsection{Without asymmetry}
Following the structure of the previous section, let us first assume that there is no asymmetry.

\begin{figure}[t]
	\centering
	\includegraphics[width=0.4\linewidth]{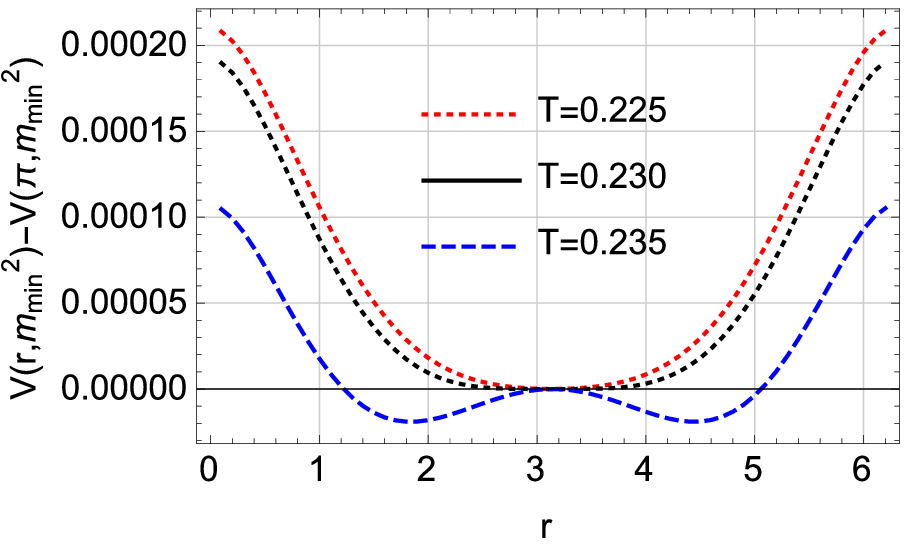}\quad\includegraphics[width=0.4\linewidth]{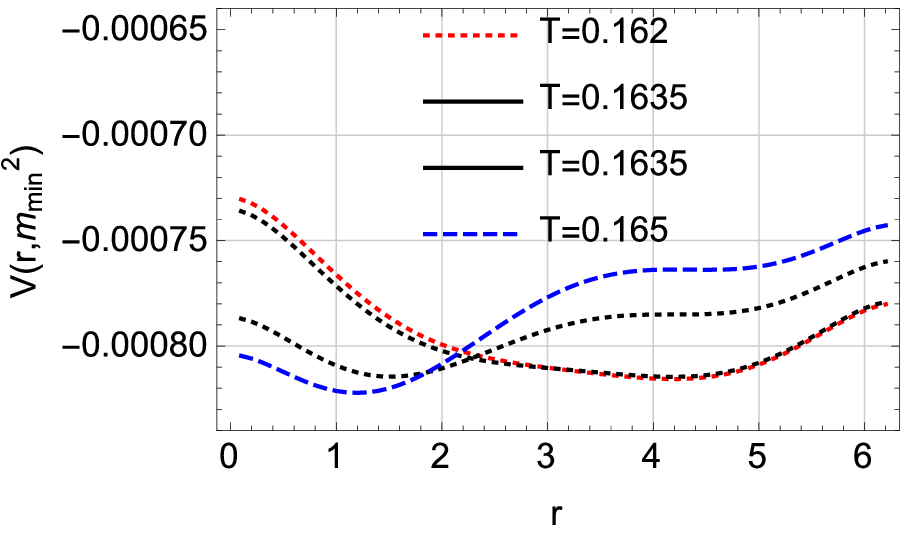}\\
		\includegraphics[width=0.4\linewidth]{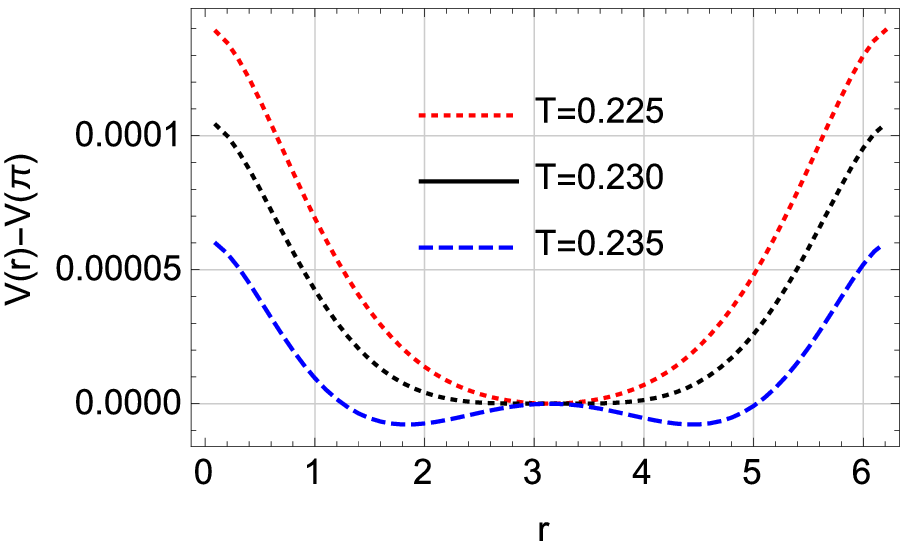}\quad\includegraphics[width=0.4\linewidth]{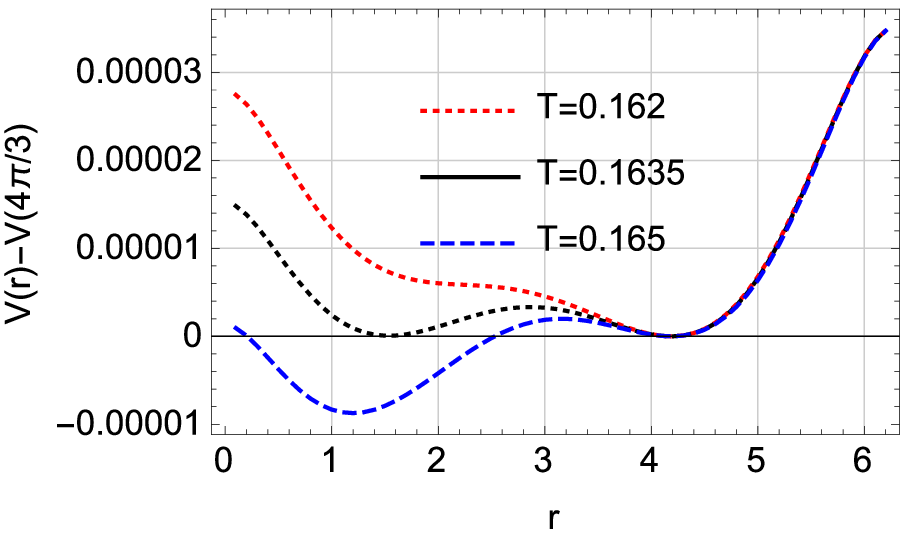}
	\caption{{Top: The potential $V(r,m^2_{\min})$ for different temperatures, in the SU(2) case (left) and in the SU(3) case (right). In this latter case, because $m^2_{\min}$ jumps at the transition, $V(r,m^2_{\min})$ does not provide the usual picture of a transition with degenerate minima. What happens is that just below $T_c$, the minimum is located at $4\pi/3$ corresponding to a certain value of $m^2_{\min}(T<T_c)$. Similarly, above $T_c$, the minimum is located away from $4\pi/3$ corresponding to a  certain value of $m^2_{\min}(T>T_c)$, and one has $m^2_{\min}(T_c^-)\neq m^2_{\min}(T_c^+)$. When approaching $T_c$ from below, $V(r,m^2_{\min}(T<T_c))$ never develops a broken symmetry form, and similarly when approaching $T_c$ from above,  $V(r,m^2_{\min}(T>T_c))$ always keeps its broken form. What happens at $T=T_c$ is that the symmetric form of $V(r,m^2_{\min}(T<T_c))$ is replaced by the broken form of $V(r,m^2_{\min}(T>T_c))$. Bottom: The reduced potential $V(r)=V(r,m^2(r))$ for different temperatures, in the SU(2) case (left) and in the SU(3) case (right). In units GeV.}}\label{fig:2}
\end{figure}

For each temperature, we find the values $r_{\min}(T)$ and $m^2_{\min}(T)$ of $r$ and $m^2$ that minimize the potential $V(r,m^2)$. We notice that the minimization might be tricky since the potential is defined only over the semi-axis $m^2\geq 0$. In particular, the absolute minimum of the potential could be located at $m^2=0$ without corresponding to a stationary point. Of course, we should follow the deepest stationary minimum as this corresponds to the limit of zero sources. The results of following this minimum are shown in Fig. \ref{fig:1}. As explained earlier, the fact that $r_{\min}$ is moving away from its center-symmetric value, $r=\pi$ in the SU(2) case and $r=4\pi/3$ in the SU(3) case, indicates deconfinement. The transition is continuous in the SU(2) case and first order in the SU(3) case as expected. This is further illustrated in Fig.~\ref{fig:2} where we show the potential as a function of $r$ for temperatures just below $T_c$, at $T_c$ and just above $T_c$. More precisely, what is shown in this figure is the potential $V(r,m^2_{\min})$ where $m^2$ has been adjusted to $m^2_{\min}$ at each temperature. Although convenient in the SU(2) case, this is not the most efficient way to illustrate the transition in the SU(3) case because $m^2_{\min}$ has a jump at $T_c$, which complicates the interpretation. This is illustrated in the right plot of Fig.~\ref{fig:2} and the corresponding caption. A more convenient quantity in this case is the reduced potential $V(r)\equiv V(r, m^2(r))$, where $m^2(r)$ is obtained by minimizing with respect to $m^2$ at fixed $r$ (by this, we mean again locating the deepest stationary minimum). The reduced potential is also shown in Fig.~\ref{fig:2}. We see that we recover the usual interpretation of the transition in the SU(3) case. As for the SU(2) case, the interpretation is essentially the same whether we use $V(r,m^2_{\min})$ or $V(r,m^2(r))$.

Using the zero-temperature parameters given above, the transition temperatures are found to be $T_c=0.230$~GeV in the SU(2) case,  and $T_c=0.1635$~GeV in the SU(3) case. Surprisingly, these values are quite close to those reported in Ref.~\cite{Reinosa:2014ooa} within the one-loop CF model. This can be understood in a simple way as follows. First, we already noticed above that the values for $m^2_{\min}$ at $T=0$ are rather close to those of the CF mass parameter, both in the SU(2) case and in the SU(3) case. Moreover, we note from Fig.~\ref{fig:1} that, below $T_c$, $m^2_{\min}$ changes marginally and, thus, remains close to its value at $T=0$. We have thus found that, at least at the present level of approximation, the dynamically generated condensate basically reproduces (from a BRST-invariant set-up) the one-loop CF model in the low temperature phase (which usually features a constant mass). It is then not a surprise that the obtained transition temperatures are close to those in the CF model.

The marginal variation of $m^2_{\min}$ with $T$ in the low temperature phase can be further understood as follows. The gap equation that determines $m^2_{\min}$ is
\begin{equation}
0=\left.\frac{\partial V}{\partial m^2}\right|_{m^2_{\min}}=\zeta m^2_{\min} \left(1-\frac{\delta\zeta}\zeta\right)+\frac{d-1}2 \tr \frac1{P^2_\kappa + m^2_{\min}}\,,
\end{equation}
where we note that only massive integrals contribute. In the low temperature phase, $r$ should be taken equal to $\pi$ (we consider the SU(2) case first for simplicity but the result generalizes to $SU(N)$). In the presence of this confining background, the tadpole integrals corresponding to $\kappa=\pm1$ become fermionic tadpole integrals. We can then separate the $T=0$ part from the thermal part and write
\begin{equation}
0=\left.\frac{\partial V_{T=0}}{\partial m^2}\right|_{m^2_{\min}}+\frac{3}{4\pi^2}\int_0^\infty dq\,\frac{q^2}{\sqrt{q^2+m^2_{\min}}} \left(\frac1{e^{\sqrt{q^2+m^2_{\min}}/T}-1}-\frac2{e^{\sqrt{q^2+m^2_{\min}}/T}+1}\right)\,.\label{eq:qq}
\end{equation}
To study  the behavior of $m^2_{\rm min}$ as $T\to 0$, we write $m^2_{\rm min}=m^2_{\rm min,T=0}+\delta m^2_{\rm min}$ with $\delta m^2_{\rm min}$ small and expand the previous formula to first non-trivial order. Because
\begin{equation}
	0=\left.\frac{\partial V_{T=0}}{\partial m^2}\right|_{m^2_{\min,T=0}}\,,
\end{equation}
this first non-trivial order yields
\begin{equation}
\delta m^2_{\min}\sim\frac{3}{4\pi^2}\frac{\int_0^\infty dq\,\frac{q^2}{\sqrt{q^2+m^2_{\min,T=0}}} \left(\frac2{e^{\sqrt{q^2+m^2_{\min,T=0}}/T}+1}-\frac1{e^{\sqrt{q^2+m^2_{\min,T=0}}/T}-1}\right)}{\left.\partial^2 V_{T=0}/\partial (m^2)^2\right|_{m^2_{\min,T=0}}}\,.
\end{equation}
The numerator can be approximated using low-temperature expansions for the tadpole integrals, whereas the denominator can be computed explicitely. We arrive at
\begin{equation}
\frac{\delta m^2_{\min}}{m_{\min,T=0}^2}\sim\frac{1}{3\pi}\left(\frac{m_{\min,T=0}}{2\pi T}\right)^{1/2}e^{-m_{\min,T=0}/T}\,,
\end{equation}
so that $m^2_{\min}$ approaches its $T=0$ value exponentially. In fact, since $m_{\min,T=0}\simeq 3T_c$, the exponential factor remains tiny over the whole confined phase, which, in turn, explains why the mass changes marginally in this phase. Above the transition, the background $r$ departs from $\pi$ and becomes a function of $T$ that introduces a new source of $T$-dependence in the right-hand side of Eq.~(\ref{eq:qq}). This explains why $m^2_{\min}$ can have a stronger variation in the deconfined phase, see Fig.~\ref{fig:1}.\\

So far, we have been concerned with the physical solution of the gap equation as given by the minimum of $V(r,m^2)$ and which corresponds to a non-zero $m^2_{\min}$ at $T=0$. At $T=0$, there is another solution, $m^2=0$, corresponding to a maximum of $V_{T=0}$. Its fate when $T>0$ is important for it controls what happens with the physical solution for $T>T_c$ as we now argue. First let us have a look at the $m^2$-derivative of $V(r_{\min},m^2)$ at $m^2=0$. This derivative reads
\begin{equation}
\left.\frac{\partial V}{\partial m^2}\right|_{m^2=0}=\frac{3}2 \tr \frac1{P^2_\kappa}=\frac{3T^2}{4}\left[B_2(0)+2B_2\left(\frac{r_{\rm min}}{2\pi}\right)\right]\,,
\end{equation}
where we have used that, in the presence of a background, the massless tadpole integral can be written in terms of the Bernouilli Polynomial $B_2(x)=x^2-x+1/6$. Now, as long as $T<T_c$, we have $r_{\min}=\pi$ (again, we consider the SU(2) case but the proof generalizes to $SU(N)$) and the term between brackets writes
\begin{equation}
B_2(0)+2B_2\left(\frac{1}{2}\right)=\frac{1}{6}+2\left(\frac{1}{4}-\frac{1}{2}+\frac{1}{6}\right)=\frac{1}{6}+2\left(-\frac{1}{12}\right)=0\,,
\end{equation} which is nothing but the well known cancellation between the bosonic and fermionic tadpole integrals appearing in (\ref{eq:qq}) in the massless case. This means that, as long as $T<T_c$, in addition to the physical minimum at $m^2_{\min}$, the potential $V(r_{\min},m^2)$ has a maximum at $m^2=0$. In contrast, whenever $T>T_c$, $r=\pi+2\pi x$ and we find
\begin{equation}
B_2(0)+2B_2\left(\frac{1}{2}+x\right)=\frac{1}{6}+2\left(\frac{1}{4}+x+x^2-\frac{1}{2}-x+\frac{1}{6}\right)=2x^2>0\,.
\end{equation}
This means that, for $T>T_c$ the maximum of $V(r_{\min},m^2)$ is pushed towards $m^2>0$ values. As we continue increasing the temperature further, we find that this maximum eventually merges with the physical minimum, thus causing the loss of both extrema above some temperature.

To clarify this feature further, we observe that, above some temperature ($T>T_1>T_c$) and below a certain background ($r<r_1(T)$), the function $V(r,m^2)$ has no extrema with respect to $m^2$. The value $r_1(T)$ corresponds to the appearance of a spinodal located at $m^2_1(T)$ at which the minimum and the maximum in $m^2$ merge. The functions $r_1(T)$ and $m^2_1(T)$ can be obtained by solving the coupled equations $\partial V/\partial m^2=0$ and $\partial^2 V/\partial (m^2)^2=0$ for each temperature above $T_1$. The function $r_1(T)$ is represented by a dotted curve in the right plots of Fig.~\ref{fig:1}. The inset in the plot shows that the curve $r_{\min}(T)$ eventually meets the curve $r_1(T)$ at a certain temperature $T_{\mbox{\tiny{loss}}}>T_1$ beyond which the solution is lost. We find $T_{\mbox{\tiny{loss}}}\simeq 0.2504$~GeV in the SU(2) case and $T_{\mbox{\tiny{loss}}}\simeq 0.1715$~GeV in the SU(3) case. In conclusion, the deconfined phase can only be explored within a tiny range of temperatures above $T_c$. Let us also mention, that the existence of $r_1(T)$ implies that, for $T>T_1$ the reduced potential introduced above is defined only for $r\in [r_1(T),2\pi-r_1(T)]$ (we consider the SU(2) case for illustration).

Finally, for completeness, we mention that there is another function $r_2(T)$ worth mentioning for $T>T_2$ (with $T_2<T_1$). For $r<r_2(T)$, the absolute minimum of $V(m^2,r)$ in the direction of $m^2$ is not a stationary point anymore (in the sense of a vanishing derivative). Rather it is located at $m^2=0$ where it has a non-vanishing derivative. This means that when $r_{\min}(T)$ goes below that line (represented by a dashed curve in the right plots of Fig.~\ref{fig:1}), we are not following the absolute minimum of the potential anymore but rather the deepest stationary minimum. This should be fine, however, since this is the point that should correspond to the limit of zero sources.\\

\begin{figure}[t]
	\centering
	\includegraphics[width=0.4\linewidth]{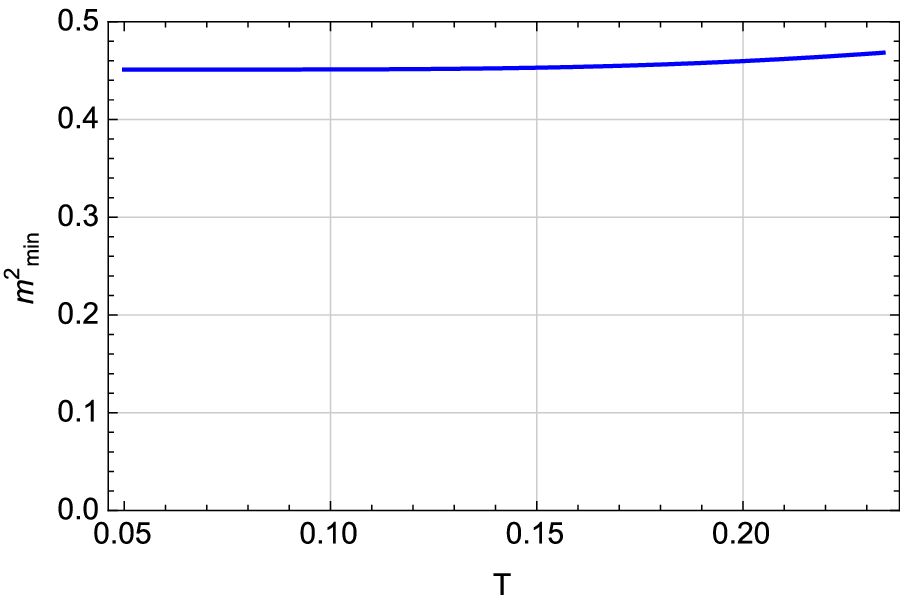}\quad\includegraphics[width=0.4\linewidth]{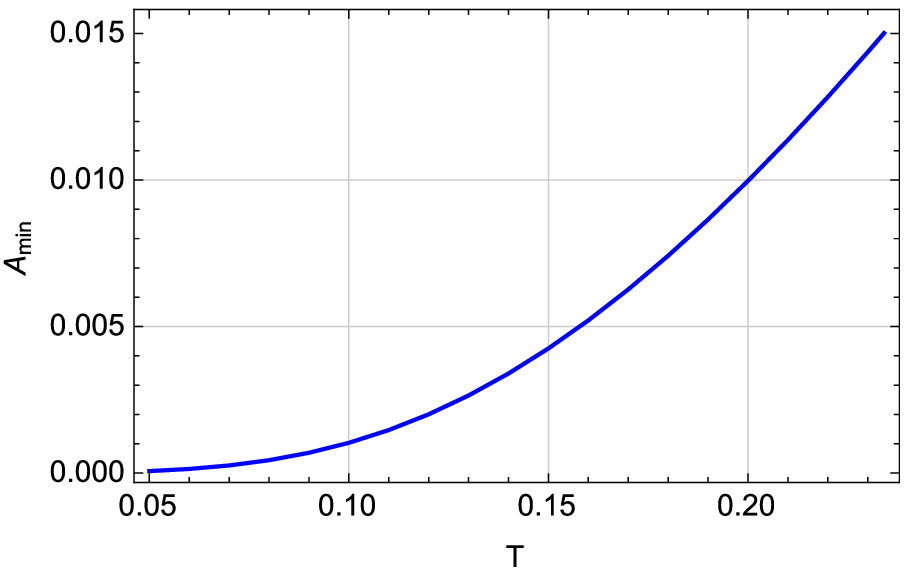}
	\caption{The parameters $m^2_{\min}$ (top) and $\mathbb{A}_{\min}$ (bottom) for the SU(2) theory, shown for temperatures where $r_{\min}=\pi$. In units GeV.}\label{fig:3}
\end{figure}

\subsection{With asymmetry}

When we allow for the asymmetry, we have to minimize the potential with respect to three parameters: $r$, $m^2$ and $\mathbb{A}$. Let us consider, for simplicity, the SU(2) case. Up to { $T=0.234$~GeV}, we find minima at $r_{\min}=\pi$ indicating that we are in the confined phase. The minimizing values for $\mathbb{A}$ and $m^2$ for $T<0.234$~GeV are given in Fig. \ref{fig:3}. Above $T=0.234$~GeV, the numerics become less trustworthy. { It should be mentioned here that the minimization is complicated by the fact that the arguments of the potential obey certain constraints. For instance, for a given $m^2$, $\mathbb{A}$ is bounded from above by $3m^2$.}

\begin{figure}[t]
 \includegraphics[width=0.45\linewidth]{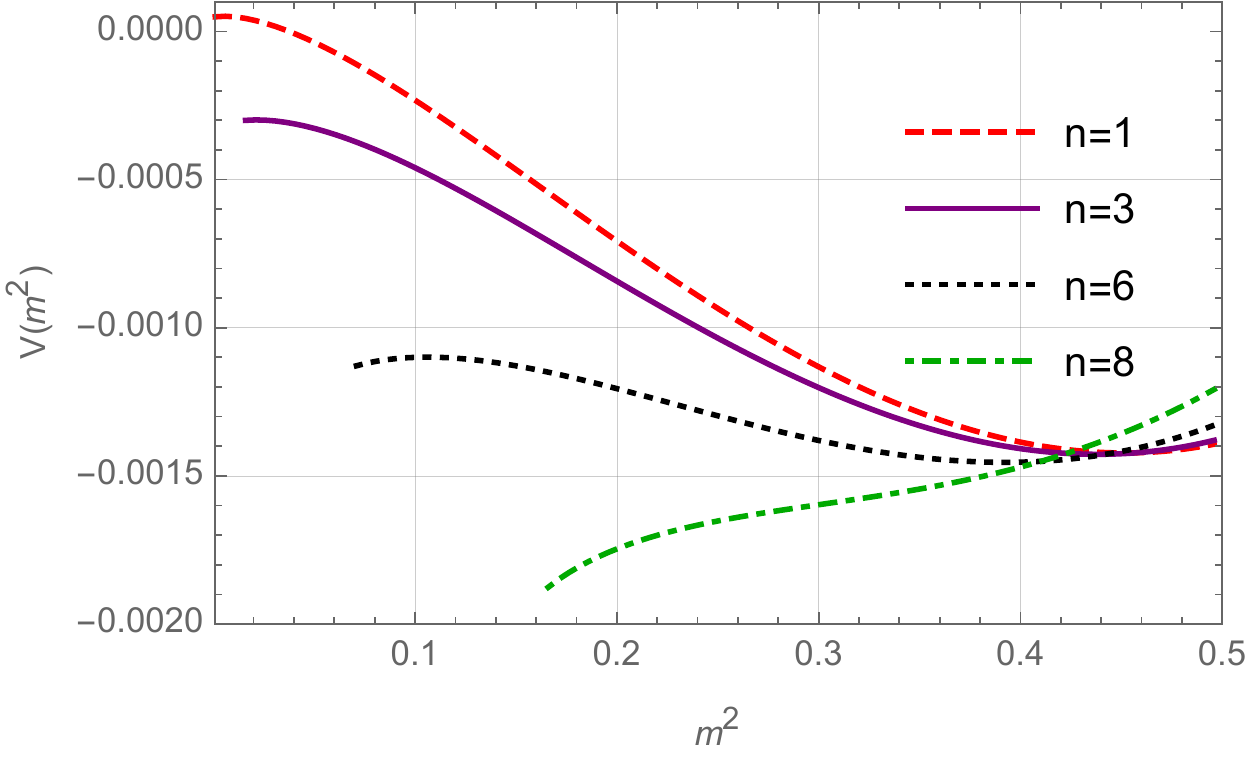}\quad \includegraphics[width=0.45\linewidth]{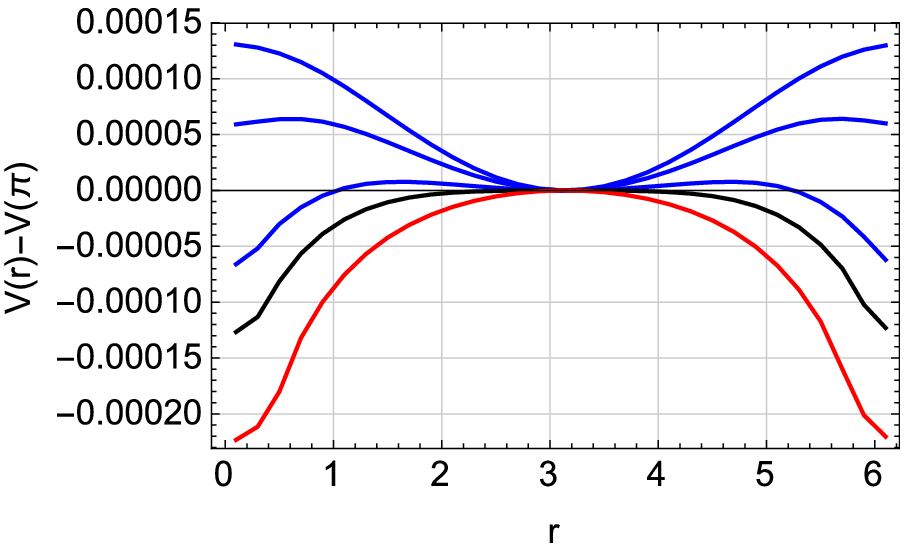}
	\caption{Left: $V(m^2)$ where  $r$ and $\mathbb{A}$ are constants found by minimizing the potential in the fine-tuning process as described in the text, given for different numbers of iterative fine-tuning $n$. Right: The reduced potential $V(r)=V(r, m^2(r), \mathbb{A}(r))$ and the loss of solution at $T_c$. All values are in GeV.  \label{iteration}}
\end{figure}

What is certain is that, above $T_c$, $r=\pi$ is no longer a minimum. However, the values for the minimizing parameters that we find numerically at $T>T_c$, taking for example $T=0.235$~GeV do not give the exact minimum of $V$. To cure this, we can in principle perform some fine-tuning in the following way: we take the values of $r$ and $\mathbb{A}$ that we found and plug them into the potential, we minimize the potential for $m^2$ and find a new value of $m^2$. We then do the same but keeping $m^2$ and $\mathbb{A}$ constant at their last-found values to find a new $r$ which minimizes the potential. Finally we do the same by keeping $r$ and $m^2$ constant and minimizing for $\mathbb{A}$. We then go back to the first step and we go on until we have found a stable minimum. As it turns out however, successively minimizing in this way leads to a loss of the stationary minimum of $V(m^2)$, as one can see in Figure \ref{iteration} where we have plotted $V(m^2)$ for several iterations of the fine-tuning procedure. It thus appears that the minimum we found for temperatures in the confining phase simply disappears above the transition. Also, the absolute minimum will now lie at $m^2=0$, where the effective potential becomes complex. Something similar was also observed in \cite{Vercauteren:2010rk}, see also \cite{Marko:2015gpa}.

The previous difficulties can be illustrated further as follows. One checks that, for any given $r$ (or at least in the vicinity of $r=\pi$), the potential admits a minimum in the plane $(m^2, \mathbb{A})$. This allows one to define $m^2(r)$ and $\mathbb{A}(r)$ and then the reduced potential $V(r)\equiv V(r, m^2(r), \mathbb{A}(r))$, similarly to what we did in the previous section. The reduced potential is shown in Fig.~\ref{iteration} for various temperatures. We see that at some temperature below $T_c$, the maximum at $r=0$ (or $r=2\pi$) starts moving into the interval $]0,2\pi[$. Eventually, it fuses with the minimum at $r=\pi$. At this spinodal, the curvature is $0$ which serves as an identification of $T_c$.\footnote{It could seem from Fig.~\ref{iteration} that, after the maximum at $r=0$ enters the $r>0$ region, there is a stationary minimum at $r=0$ that can become the absolute one before the destabilization of the one at $r=\pi$, what would then as a first order transition. We have checked, however, that, before this happens, the reduced potential in the vicinity of $r=0$ is not well defined because of the inability to find a stationary $(m(r), \mathbb{A}(r))$.} At the same time, however, the minimum disappears and one cannot follow it into the deconfined phase. Our conclusion is then that, in the presence of asymmetry, we have no access to the deconfined phase, at least not within our current level of treatment.\\

It is commonly known that at high(er) temperatures, resummations are in order to save the perturbative expansion, see e.g.~\cite{Braaten:1989mz,Chiku:1998kd} or the reviews \cite{Andersen:2004fp,Laine:2016hma}. We will not attempt this here, as our main focus was on determining the deconfinement transition and its interplay with the dimension two condensates. Our findings are however clear: the critical deconfinement estimate, $T_c\approx0.231$ GeV, is very close to the one loop estimate reported in \cite{Reinosa:2014ooa} using the $T=0$ Curci--Ferrari mass fit parameter, namely $T_c\approx0.238$~GeV. We repeat here we did not use any external (lattice) input, except for the estimate of $\lms$ of course. A posteriori, this is not such a surprise: the used value for the (temperature independent) gluon mass in \cite{Reinosa:2014ooa} was $m(T)\approx0.710$~GeV, whilst here we find that the dynamical $m(T)$ indeed varies only little from its $m(T=0)\approx0.670$~GeV value (cfr.~\eqref{massa}), next to a pretty small asymmetry. For the record, functional approaches as in \cite{Fister:2013bh} arrived at $T_c=0.230\pm0.023$~GeV, \textit{i.e.} all values in the same ballpark. This extends to the variational estimate of \cite{Quandt:2016ykm}, $T_c\approx0.239$~GeV, or the Coulomb gauge variational estimate $T_c\approx0.275-0.290$~GeV \cite{Heffner:2012sx}. For comparison, lattice estimates for the SU(2) transition temperate are $T_c\approx0.295$~GeV \cite{Lucini:2012gg} or $T_c\approx0.312$~GeV \cite{Lucini:2003zr,Quandt:2016ykm}.

We find similar difficulties for the SU(3) case.

Let us spend a few more words on the asymmetry $\mathbb{A}$. Next to the pioneering Landau gauge lattice study of \cite{Chernodub:2008kf} and later efforts as in \cite{Bornyakov:2016geh,Bornyakov:2022wxc}, the only analytical investigation of it so far is \cite{Vercauteren:2010rk} by one of us, see also \cite{Dudal:2009tq}. As the Polyakov loop was not part of that approach, nothing could be said about the sensitivity of the asymmetry to the phase transition. Now we do have such evidence by working in the Landau--DeWitt gauge, albeit that the findings are not exactly numerically comparable, not only because we do not have results in the deconfined phase, but also the magnitude of the asymmetry is considerably smaller than that reported on the lattice.

\section{Conclusions}
Let us end by pointing out a few possible routes towards a further development of the framework here proposed. It has been conjectured that the electric and magnetic {Landau gauge} propagators at zero momentum, corresponding to the respective screening masses, are sensitive to the phase transition (and even its order) in \cite{Maas:2011ez}.\footnote{Relatedly, the integrated difference between electric and magnetic propagator in relation to the transition was studied recently in \cite{Bornyakov:2022wxc}.} {This scenario is however far from clear in the Landau gauge \cite{Cucchieri:2011tlu,Silva:2013maa,Silva:2016onh}. In fact, it has been argued in \cite{vanEgmond:2021jyx,vanEgmond:2022nuo} that there is no actual reason for the Landau gauge propagators to be sensitive to the transition because the average gluon field is not an order parameter in this case. Still in \cite{vanEgmond:2021jyx,vanEgmond:2022nuo}, a particular background Landau gauge (referred to as {\it center-symmetric}) has been put forward in which the background takes a center-invariant configuration in any phase. In this gauge, the average gluon field becomes an order parameter and, correspondingly, the electric propagator shows distinctive features at the transition \cite{vanEgmond:2022nuo}. It would be interesting to revisit the present considerations in this particular gauge. Evidently, one expects the asymmetric gluon condensate to influence exactly the aforementioned quantities.

The center-symmetric Landau gauge is closely related to the formalism used in the present work and based on the background effective potential. However, if the former relies on the use of a standard Legendre transform, the justification of the latter (and the equivalence with the former) relies on the background independence of the free energy \cite{Reinosa:2020mnx}. This property is not necessarily met identically within an approximated setting or in the presence of modelling.\footnote{By modelling, we refer to possible Ans\"atze for the vertex functions in DSE/FRG approaches, the phenomenological addition of operators beyond the incomplete Faddeev--Popov action (as is done for instance within the Curci--Ferrari model), etc.} For instance, in the case of the Curci--Ferrari model, the use of the center-symmetric Legendre transform leads to improved predictions as compared to those obtained using the background effective potential. In the present, BRST based approach, the background independence of the observables, and in particular of the free energy, is protected by the combined use of  the BRST symmetry of the action, the BRST exactness of the Faddeev--Popov action, and the fact that the background dependence in the $\sigma$-sector cancels identically upon exact integration of $\sigma$. So, even though, as we have seen, the dynamical condensate mimics the Curci--Ferrari model in the low temperature phase, we expect a better account of the background independence of the observables and therefore a better agreement between the present approach and that based on the center-symmetric Landau gauge. Results in this direction will be reported elsewhere.}

Overall, the results presented are eventually rather similar to the ones of the phenomenological massive Curci--Ferrari--Landau--DeWitt model \cite{Reinosa:2014ooa,Reinosa:2014zta,Pelaez:2021tpq}, the big step forward being that the (crucial) non-perturbative mass scales now have a dynamical origin and that BRST is maintained. A posteriori, our setup grants credit to the aforementioned model, even at the quantitative level as we have discussed. At least at one-loop order, we do not expect much will change for what concerns the other thermodynamical observables such as pressure, entropy, trace anomaly, etc. when compared to \cite{Reinosa:2014zta}, but a more thorough discussion of this is relegated to future work. Furthermore, although our results are non-perturbative in nature, they do arise from an effective potential computed in a loop expansion. It remains to be investigated how stable the results are against adding the two-loop corrections, which should still be analytically tractable since the most complicated pieces have already been computed \cite{Reinosa:2014zta}. Notice that from two-loop onwards, more changes might occur relative to the Curci--Ferrari--Landau--DeWitt model, as the other vertices containing the $\sigma$- and $\varphi$- fields (arising from expanding the action of the unity \eqref{uniteit} around the would-be condensates), will enter the game.

\section*{Acknowledgments}
D.~Dudal acknowledges financial support from Ecole Polytechnique (Institut Polytechnique de Paris) and CNRS, next to the warm hospitality at CPHT, where parts of this work were prepared. D.M.~van Egmond was partly financed by KU Leuven with a visiting researcher fellowship. D.~Vercauteren is grateful for the hospitality at KU Leuven, made possible through the Senior Fellowship SF/19/008. We thank A.D.~Pereira and G.~Comitini for interesting discussions.

\appendix
\section{Renormalizability of dimension two local composite operators in the background gauge} \label{backgroundren}
In this Appendix, we show the renormalizability to all orders in an algebraic setting of the LCO formalism in the presence of a classical gauge background field $\bar A_\mu^a$:
\begin{equation}
a_\mu^a = A_\mu^a + \bar A_\mu^a \ ,
\end{equation}
where $A_\mu^a$ represents the quantum part of the gauge field. The Landau gauge condition $\partial_\mu A_\mu^a=0$ is now replaced by\footnote{Mark that, if $\bar A_\mu^a$ is chosen to satisfy the Landau gauge, many of the expressions in this section will simplify considerably. Nevertheless, even if one plans to choose the Landau gauge for the quantum fluctuations, it is for now more opportune to leave $\bar A_\mu^a$ more general, as it will allow to take functional derivatives with respect to $\bar A_\mu^a$ and $\Omega_\mu^a$ more freely.}
\begin{equation}
\bar D_\mu A_\mu^a = \bar D_\mu(a_\mu^a-\bar A_\mu^a) = 0 \ ,
\end{equation}
where $\bar D_\mu$ is the covariant derivative containing only the background field $\bar A_\mu^a$. In this gauge the ghost action is changed accordingly to
\begin{equation}
\mathcal L_\text{gh} = \int d^4x \; \bar c^a\bar D_\mu D_\mu c^a \ .
\end{equation}
The condensate $\langle A_\mu^2\rangle$ we want to compute the vacuum expectation value of will, of course, also need to be modified. It turns out that, if we demand renormalizability of the action, the operator $A_\mu^2 = (a_\mu^a-\bar A_\mu^a)^2$ is to be considered. Let us now prove that this is indeed the only possible choice. For this we use the algebraic renormalization formalism, and the computations outlined below are parallel to those done by one of us and coworkers in the linear covariant gauges \cite{brst}. The algebraic analysis of the background gauge has already been explored in \cite{brstbackground}, and their approach is used in the following.

\subsection{The classical action}
We start from the action of pure Yang--Mills theory in the Landau background gauge:
\begin{equation}
S_\text{YM+gf} = \int d^4x \left(\frac14(F_{\mu\nu}^a)^2 + ib^a\bar D_\mu(a_\mu^a-\bar A_\mu^a) + \bar c^a\bar D_\mu D_\mu c^a\right) \ .
\end{equation}
Here we introduced the Nakanishi--Lautrup field $b^a$, which is a Lagrange multiplier for the gauge fixing condition. A second part of the action consists of the source field $J$ coupled to the operator we are considering, which we leave more general for now:
\begin{equation}
S_J = \int d^4x \left(\frac12J(a_\mu^2 + \alpha a_\mu^a\bar A_\mu^a + \beta\bar A_\mu^2) + \frac\zeta2 J^2\right) \ ,
\end{equation}
where the term in $J^2$ has been added to absorb the quadratic divergences in the source field. The numbers $\alpha$ and $\beta$ will be determined by demanding renormalizability. The parameter $\zeta$ has to be introduced here, and, just as in the case without a background field, it will have to be determined using other considerations, cfr.~the main body of this paper. Finally we introduce classical source fields $\Delta^*$, ${A^*}_\mu^a$, and ${c^*}^a$ coupling to the nonlinear BRST variations of the fields and operators under consideration:
\begin{equation}
S_\text{ext} = \int d^4x \Bigg(\Delta^*\left((a_\mu^a+\tfrac\alpha2\bar A_\mu^a)D_\mu c^a - (\tfrac\alpha2a_\mu^a+\beta\bar A_\mu^a)\Omega_\mu^a\right) - {A^*}_\mu^a(D_\mu c^a+\Omega_\mu^a) + \frac12gf^{abc}{c^*}^ac^bc^c\Bigg) \ ,
\end{equation}
where we have introduced the ghost field $\Omega_\mu^a$, which is the BRST transformation of $\bar A_\mu^a$. The term in ${A^*}_\mu^a\Omega_\mu^a$ has been added in order to allow to absorb the counterterms later on. If we add one final term
\begin{equation}
\int d^4x \; \bar c^aD_\mu\Omega_\mu^a
\end{equation}
necessary to cancel some spurious terms coming from the BRST variation of $\bar A_\mu^a$, the total action is invariant under the nilpotent BRST transformation $s$ defined by
\begin{gather}
sa_\mu^a = -D_\mu c^a \ , \qquad sc^a = \frac12gf^{abc}c^bc^c \ , \nonumber \\
s\bar c^a = ib^a \ , \qquad s\bar A_\mu^a = \Omega_\mu^a \ , \qquad s\Delta^* = J \ , \\
sb^a = sJ = s\Omega_\mu^a = s{A^*}_\mu^a = s{c^*}^a = 0 \ . \nonumber
\end{gather}
Notice that the possibility of introducing the background field $\bar A_\mu$ as part of a BRST doublet is almost immediately leading to the independence on $\bar A_\mu$ of observables, defined as elements of the BRST cohomology with zero ghost charge \cite{Piguet:1995er,Ferrari:2000yp}. Indeed, doublets are trivial elements of the cohomology \cite{Piguet:1995er}.

The full action can be rewritten in the form
\begin{equation}
	S = \frac14\int d^4x(F_{\mu\nu}^a)^2 + s\int d^4x\Bigg(\bar c^a\bar D_\mu(a_\mu^a-\bar A_\mu^a) + \frac12\Delta^*(a_\mu^2 + \alpha a_\mu^a\bar A_\mu^a + \beta\bar A_\mu^2) - {A^*}_\mu^a(a_\mu^a-\bar A_\mu^a) + {c^*}^a c^a + \frac\zeta2\Delta^*J\Bigg) \ .
\end{equation}
From this form, the BRST invariance is easy to see: working with $s$ on the first term will give a mere gauge transformation with $c^a$ as the gauge function, and working on the second part will give zero as $s$ is nilpotent by definition.

At the classical level, the theory is characterized by some powerful identities. We have the Slavnov--Taylor identity:
\begin{subequations} \label{drievergelijkingen} \begin{equation}
\mathcal S(S) = \int d^4x \left(\frac{\delta S}{\delta a_\mu^a} \frac{\delta S}{\delta{A^*}_\mu^a} + \frac{\delta S}{\delta c^a} \frac{\delta S}{\delta{c^*}^a} + ib^a\frac{\delta S}{\delta\bar c^a} + \Omega_\mu^a \frac{\delta S}{\delta\bar A_\mu^a} + J\frac{\delta S}{\delta\Delta^*}\right) = 0 \ ,
\end{equation}
which is nothing but a reexpression of the BRST invariance of the action; the equation for the Nakanishi--Lautrup field:
\begin{equation}
\frac{\delta S}{\delta b^a} = i\bar D_\mu(a_\mu^a-\bar A_\mu^a) \;,
\end{equation}
the antighost equation:
\begin{equation}
\frac{\delta S}{\delta\bar c^a} + \bar D_\mu \frac{\delta S}{\delta{A^*}_\mu^a} = D_\mu\Omega_\mu^a \ ,
\end{equation}
which can be straightforwardly found by taking the derivative with respect to the antighost field $\bar c^a$ and rewriting the composite operator $D_\mu c^a$ as a derivative with respect to ${A^*}_\mu^a$; and the ghost Ward identity:
\begin{equation} \label{spookvgl}
\frac{\delta S}{\delta c^a} + \bar D_\mu\frac{\delta S}{\delta\Omega_\mu^a} - igf^{abc}\bar c^b\frac{\delta S}{\delta b^c} =  (1+\tfrac\alpha2)\partial_\mu(\Delta^*a_\mu^a) + (\tfrac\alpha2+\beta)\partial_\mu(\Delta^*\bar A_\mu^a) - D_\mu {A^*}_\mu^a + gf^{abc}c^b{c^*}^c \ .
\end{equation} \end{subequations}
This last identity can be found by first taking the derivative of the action with respect to the ghost field $c^a$:
\begin{equation}
\frac{\delta S}{\delta c^a} = -D_\mu\bar D_\mu\bar c^a + \partial_\mu(\Delta^*a_\mu^a) + \tfrac\alpha2D_\mu(\Delta^*\bar A_\mu^a) - D_\mu {A^*}_\mu^a + gf^{abc}c^b{c^*}^c \ .
\end{equation}
Then we note that
\begin{equation}
[\bar D_\mu,D_\mu]^{ab} = -gf^{abc}\bar D_\mu(a_\mu^c-\bar A_\mu^c) = igf^{abc} \frac{\delta S}{\delta b^c} \ ,
\end{equation}
which gives
\begin{equation}
\frac{\delta S}{\delta c^a} - igf^{abc}\bar c^b\frac{\delta S}{\delta b^c} = -\bar D_\mu D_\mu\bar c^a + \partial_\mu(\Delta^*a_\mu^a) + \tfrac\alpha2D_\mu(\Delta^*\bar A_\mu^a) - D_\mu {A^*}_\mu^a + gf^{abc}c^b{c^*}^c \ .
\end{equation}
In order to get rid of the composite operator term with $D_\mu\bar c^a$, we consider:
\begin{equation}
\frac{\delta S}{\delta\Omega_\mu^a} = \tfrac\alpha2\Delta^*a_\mu^a + \beta\Delta^*\bar A_\mu^a + D_\mu\bar c^a \ .
\end{equation}
Using this, we immediately find \eqref{spookvgl}.

\subsection{The most general counterterm}
When doing perturbation theory, counterterms have to be added to the classical theory. If we write this as $S+\epsilon S^\text{ct}$, where $\epsilon$ is the perturbation parameter, then we can demand the full action to obey the same set of identities \eqref{drievergelijkingen} up to leading order in $\epsilon$. For the counterterm, this translates to the conditions:
\begin{subequations} \begin{gather}
\mathcal B_S S^\text{ct} = 0 \ , \label{slavnovtaylor} \\
\frac{\delta S^\text{ct}}{\delta b^a} = 0 \ , \label{nakanishilautrup} \\
\frac{\delta S^\text{ct}}{\delta\bar c^a} + \bar D_\mu\frac{\delta S^\text{ct}}{\delta{A^*}_\mu^a} = 0 \label{antispook} \ , \\
\frac{\delta S^\text{ct}}{\delta c^a} + \bar D_\mu\frac{\delta S^\text{ct}}{\delta\Omega_\mu^a} = 0 \label{spookvglct} \ ,
\end{gather}
where $\mathcal B_S$ is the linearized operator:
\begin{equation}
\mathcal B_S = \int d^4x \Bigg(\frac{\delta S}{\delta a_\mu^a}\frac\delta{\delta{A^*}_\mu^a} + \frac{\delta S}{\delta{A^*}_\mu^a}\frac\delta{\delta a_\mu^a} + \frac{\delta S}{\delta c^a}\frac\delta{\delta{c^*}^a} + \frac{\delta S}{\delta{c^*}^a}\frac\delta{\delta c^a} + ib^a\frac\delta{\delta\bar c^a} + \Omega_\mu^a\frac\delta{\delta\bar A_\mu^a} + J\frac\delta{\delta\Delta^*}\Bigg) \ ,
\end{equation} \end{subequations}
which is again nilpotent. Now it follows from the general theory concerning algebraic renormalization that the most general invariant local counterterm can be parameterized as
\begin{equation}
S^\text{ct} = \frac{p_1}4\int d^4x(F_{\mu\nu}^a)^2 + \mathcal B_S\int d^4x \; \Xi \ ,
\end{equation}
where $\Xi$ is the most general local polynomial with dimension 4 and ghost number $-1$. In order to write this down, we need the dimensions and ghost numbers of the fields and sources:
\begin{center}
\begin{tabular}{|l|*{10}c|}
\hline
 & $a_\mu^a$ & $\bar A_\mu^a$ & $c^a$ & $\bar c^a$ & $b^a$ & $J$ & $\Omega_\mu^a$ & ${A^*}_\mu^a$ & ${c^*}^a$ & $\Delta^*$ \\
\hline
dimension & $1$ & $1$ & $0$ & $2$ & $2$ & $2$ & $1$ & $3$ & $4$ & $2$ \\
ghost number & $0$ & $0$ & $1$ & $-1$ & $0$ & $0$ & $1$ & $-1$ & $-2$ & $-1$ \\
\hline
\end{tabular}
\end{center}
With this we can write down the most general form for $\Xi$:
\begin{multline}
\Xi = p_2a_\mu^a{A^*}_\mu^a + p_3\bar A_\mu^a{A^*}_\mu^a + p_4c^a{c^*}^a + p_5a_\mu^a\partial_\mu\bar c^a + p_6\bar A_\mu^a\partial_\mu\bar c^a + p_7gf^{abc}\bar A_\mu^aa_\mu^b\bar c^c + p_8gf^{abc}\bar c^a\bar c^bc^c + p_9b^a\bar c^a \\ + p_{10}\Delta^*a_\mu^2 + p_{11}\Delta^*\bar A_\mu^aa_\mu^a + p_{12}\Delta^*\bar A_\mu^2 + p_{13}\Delta^*\bar c^ac^a + p_{14}\Delta^*J \ .
\end{multline}
The $p_i$, $i=1,\ldots,14$, are arbitary parameters. With this form, the constraint \eqref{slavnovtaylor} is automatically fulfilled. The equation \eqref{nakanishilautrup} gives:
\begin{equation}
	i\bar D_\mu(p_2a_\mu^a+p_3\bar A_\mu^a) - ip_5\partial_\mu a_\mu^a - ip_6\partial_\mu\bar A_\mu^a + ip_7gf^{abc}\bar A_\mu^ba_\mu^c + 2ip_8gf^{abc}\bar c^bc^c + 2ip_9b^a - ip_{13}\Delta^*c^a = 0 \ ,
\end{equation}
from which we find
\begin{equation}
	p_2 = p_5 = -p_7 \ , \qquad p_3 = p_6 \ , \qquad p_8 = p_9 = p_{13} = 0 \ .
\end{equation}
The constraint \eqref{antispook} is now already satisfied. The ghost Ward identity \eqref{spookvglct} gives:
\begin{multline}
(p_2+p_3)\partial_\mu\bar D_\mu\bar c^a + p_4D_\mu\bar D_\mu\bar c^a - p_4gf^{abc}c^b{c^*}^c + (p_2+p_3+p_4)\partial_\mu{A^*}_\mu^a + p_4D_\mu{A^*}_\mu^a \\ + \left(p_2(1+\tfrac\alpha2)-p_4+2p_{10}+p_{11}\right)\partial_\mu(\Delta^*a_\mu^a) + \left(p_3(1+\tfrac\alpha2)-p_4\tfrac\alpha2+p_{11}+2p_{12}\right)\partial_\mu(\Delta^*\bar A_\mu^a) + p_4\tfrac\alpha2gf^{abc}\Delta^*\bar A_\mu^ba_\mu^c = 0 \ ,
\end{multline}
which yields
\begin{equation}
p_2=-p_3 \ , \qquad p_{11} = -p_2(1+\tfrac\alpha2)-2p_{10} \ , \qquad p_{12} = p_2(1+\tfrac\alpha2)+p_{10} \ , \qquad p_4 = 0 \ .
\end{equation}
Finally we find for $\Xi$:
\begin{equation} \label{xigedaan}
\Xi = p_2(a_\mu^a-\bar A_\mu^a)\left({A^*}_\mu^a+\bar D_\mu\bar c^a -(1+\tfrac\alpha2)\Delta^*\bar A_\mu^a\right) + p_{10}\Delta^*(a_\mu^a-\bar A_\mu^a)^2 + p_{14}\Delta^*J \ .
\end{equation}

\subsection{Absorbing the counterterm}
From equation \eqref{xigedaan}, we can write down the most general counterterm consistent with the symmetries of the theory:
\begin{multline}
S^\text{ct} = \int d^4x\Bigg((\tfrac{p_1}4-p_2)(F_{\mu\nu}^a)^2 + p_2(\partial_\mu a_\nu^a)F_{\mu\nu}^a + p_2(D_\mu\bar A_\nu^a)F_{\mu\nu}^a - p_2\bar c^a\bar D^2c^a + (p_2+p_{10})J(a_\mu^a-\bar A_\mu^a)^2 + p_{14}J^2 \\ + \Delta^*(a_\mu^a-\bar A_\mu^a)\Big((p_2+2p_{10})\partial_\mu c^a + (p_2\tfrac\alpha2-2p_{10})gf^{abc}c^b\bar A_\mu^c + (-p_2\tfrac\alpha2+2p_{10})\Omega_\mu^a\Big) \\ - p_2(1+\tfrac\alpha2)\Delta^*(\bar A_\mu^aD_\mu c^a-\Omega_\mu^a(a_\mu^a-2\bar A_\mu^a)) + p_2{A^*}_\mu^a(\bar D_\mu c^a+\Omega_\mu^a) - p_2\Omega_\mu^a\bar D_\mu\bar c^a\Bigg) \ .
\end{multline}
Now it is clear that, in order to reabsorb this counterterm into the classical action, we need to have $\alpha=-2$ and $\beta=1$. Then we can absorb the counterterm with multiplicative renormalization. If we write the bare fields as $\Phi_0 = Z_\Phi^{1/2}\Phi$ for the fields $A_\mu^a = a_\mu^a-\bar A_\mu^a$, $c^a$, $\bar c^a$, and $b^a$, then we find:
\begin{subequations} \begin{equation}
Z_{A}^{1/2} = Z_b^{-1/2} = 1+\epsilon\left(\frac{p_1}2-p_2\right) \ , \qquad Z_c^{1/2} = Z_{\bar c}^{1/2} = 1-\epsilon\frac{p_2}2 \ .
\end{equation}
For the parameters we write $g_0 = Z_gg$ and $\zeta_0 = Z_\zeta\zeta$, and we find:
\begin{equation}
Z_g = 1-\epsilon\frac{p_1}2 \ , \qquad Z_\zeta = 1 + \epsilon\left(2p_1-4p_{10}+\frac2\zeta p_{14}\right) \ .
\end{equation}
For the sources $J$, $\Delta^*$, ${A^*}_\mu^a$, and ${c^*}^a$ we write $\Phi_0 = Z_\Phi\Phi$:
\begin{equation} \begin{gathered}
Z_J = 1 + \epsilon(-p_1+2p_{10}) \ , \qquad Z_{\Delta^*} = 1 + \epsilon\left(-\frac{p_1}2 + \frac{p_2}2 + 2p_{10}\right) \ , \\ Z_{A^*} = Z_c^{1/2} \ , \qquad Z_{c^*} = Z_{A}^{1/2} \ .
\end{gathered} \end{equation}
For the classical fields $\bar A_\mu^a$ and $\Omega_\mu^a$, we write the bare fields as $\Phi_0 = Z_\Phi^{1/2}\Phi$, and we find:
\begin{equation}
Z_{\bar A}^{1/2} = Z_g^{-1} \ , \qquad Z_\Omega = Z_c^{-1/2} \ .
\end{equation} \end{subequations}
We mark that $A_\mu^a$ and $\bar A_\mu^a$ renormalize separately. For this reason one must consider the local composite operator $A_\mu^2:=(a_\mu-\bar A_\mu)^2$ instead of $a_\mu^2$, which would not be multiplicatively renormalizable.

As we are working in a different gauge, one could expect the $\zeta$ parameter to be modified. However, this will not be the case for dimensional reasons. In the limit $\bar A_\mu^a\rightarrow0$, the Landau background gauge reduces to the ordinary Landau gauge, and so the value for $\zeta$ should be equal to the backgroundless value in that limit. Introducing a background field cannot modify it, as there are no other dimensionful quantities present to make a dimensionless function.\footnote{We work in mass-independent renormalization schemes.} This argument also carries through for the renormalization group parameters. We can conclude that the values in equations \eqref{zetadeltazeta} are valid in the Landau background gauge as well.

A final interesting few words can be said about the special values $\alpha=-2$, $\beta=1$. These are also the unique values for which the action enjoys an extra Ward identity, namely
\begin{equation}\label{backgroundwarid}
  \left(-D_\mu^{ab} \frac{\delta}{\delta a_ \mu^b}-\bar D_\mu^{ab}\frac{\delta}{\delta {\bar A}_ \mu^b}-\sum_{\Phi}gf^{abc}\Phi^b \frac{\delta}{\delta \Phi^c}\right)S=0\;,
\end{equation}
where $\Phi$ runs over all other fields/sources with an adjoint color index. This identity encodes nothing else than the background gauge invariance. The quantum stability of the specific dimension two operator $A_\mu^2$ can thus also be appreciated from background gauge invariance. For the record, as noted in \cite{brstbackground}, the identity \eqref{backgroundwarid} follows from the anti-commutator of the ghost equation \eqref{spookvgl} and the Slavnov-Taylor identity \eqref{drievergelijkingen}, at least for the identified values of $\alpha$, $\beta$.

\subsection{Inclusion of the asymmetry}
We are skipping details here, as the discussion is very similar to the one of \cite{Dudal:2009tq}. In a nutshell, once the renormalizability of the theory with coupling to it of $(a_\mu-\bar A_\mu)^ 2$ is handled, the introduction of another BRST doublet of sources, $s\eta_{\mu\nu}=K_{\mu\nu}$, $sK_{\mu\nu}=0$ allows to couple (the traceless part of) $(a_\mu-\bar A_\mu)(a_\nu-\bar A_\nu)$ to the theory in a BRST invariant fashion without hampering the other identities, leading yet again to the quantum stability upon inclusion of a pure vacuum term quadratic in the new source $K_{\mu\nu}$. Just as in \cite{Dudal:2009tq}, there will be no mixing between the two sources $J$ and $K_{\mu\nu}$.
Intuitively, renormalizability is expected, as at $T=0$, one does not expect a non-vanishing asymmetry, while no new UV divergences should emerge at $T>0$.

\section{Evaluation of the potential from the Nakanishi--Lautrup framework} \label{NL}
In the Nakanishi-Lautrup framework, one implements the $\alpha\to 0$ limit explicitly at the level of the action. The quadratic part reads
\begin{eqnarray}
	&&\int d^dx \frac{1}{2}\Bigg(A_\mu^a \left(- \delta_{\mu\nu} \bar D^2_{ab} + \bar D_\nu^{ac} \bar D_\mu^{cb} + \delta^{ab}\delta_{\mu\nu} m^2 + \delta^{ab}M_{\mu\nu} \right) A_\nu^b+\bar c^a \bar D^2_{ab} c^b +ib^a\bar D_\mu^{ab}A_\mu^b\Bigg) \;,
\end{eqnarray}
where $h^a$ is the Nakanishi-Lautrup field. In Fourier space, and in the color diagonal basis, in the $A,h$ sector, this corresponds to the matrix
\begin{eqnarray}
\left(\begin{array}{cc}
(P^2_\kappa+m^2)\delta_{\mu\nu}+M_{\mu\nu}-P^\kappa_\mu P^\kappa_\nu & P^\kappa_\mu\\
-P^\kappa_\nu & 0
\end{array}\right)\,.
\end{eqnarray}
In order to evaluate the determinant, we can always choose a frame where $P^\kappa_1=|\vec{p}|$ and $P^\kappa_{\mu>1}=0$. We find
\begin{eqnarray}
\det\left(\begin{array}{cccccc}
p^2+m^2+\mathbb{A} & -P^\kappa_0 p & 0 & & 0 & P^\kappa_0\\
-P^\kappa_0 p & (P^\kappa_0)^2+m^2-\frac{\mathbb{A}}{d-1} & 0 & & 0 & p\\
0 & 0 & P^2_\kappa+m^2-\frac{\mathbb{A}}{d-1} & & 0 & 0\\
 &  & & \ddots &  & \\
0 & 0 & 0 & & P^2_\kappa+m^2-\frac{\mathbb{A}}{d-1} & 0\\
-P^\kappa_0 & -p & 0 & & 0 & 0
\end{array}\right)\,.
\end{eqnarray}
Upon exchanging the third and last line and the third and last column, this becomes
\begin{eqnarray}
\det\left(\begin{array}{cccccc}
p^2+m^2+\mathbb{A} & -P^\kappa_0 p & P^\kappa_0 & 0 & & 0\\
-P^\kappa_0 p & (P^\kappa_0)^2+m^2-\frac{\mathbb{A}}{d-1} & p & 0 & & 0\\
-P^\kappa_0 & -p & 0 & 0 & & 0\\
0 & 0 & 0 & P^2_\kappa+m^2-\frac{\mathbb{A}}{d-1} & & 0\\
 &  & & & \ddots &\\
0 & 0 & 0 & 0 & & P^2_\kappa+m^2-\frac{\mathbb{A}}{d-1}
\end{array}\right)\,,
\end{eqnarray}
which is then easily computed to be
\begin{eqnarray}
& & \left(P^2_\kappa+m^2-\frac{\mathbb{A}}{d-1}\right)^{d-2}\left\{(P^\kappa_0)^2\left(P^2_\kappa+m^2-\frac{\mathbb{A}}{d-1}\right)+p^2(P^2_\kappa+m^2+\mathbb{A})\right\}\nonumber\\
& & \hspace{0.5cm}=\,\left(P^2_\kappa+m^2-\frac{\mathbb{A}}{d-1}\right)^{d-2}\left\{P_\kappa^2(P^2_\kappa+m^2)+\mathbb{A}\left(P^2_\kappa-\frac{d}{d-1}(P^\kappa_0)^2\right)\right\},
\end{eqnarray}
which leads to Eq.~(\ref{eq:pot_A}) upon inclusion of the ghost contribution.

\section{Sums at finite temperature}\label{sums}
At finite temperature, the imaginary time dimension is compactified with a circumference of $1/T$. This results in a discretization of the spectrum. In order to compute traces, the following replacement has to be made for bosons (including the (anti-)ghosts)
\begin{equation}
\int \frac{dk_0}{2\pi} f(k_0) \rightarrow T \sum_{n=-\infty}^{+\infty} f(2\pi nT) \;.
\end{equation}
In order to do computations, we will also need to compute sums of particle propagators. An example from which we can derive the formulae necessary in the main text is
\begin{equation} \label{eindigetvoorbeeldsom}
\sum_{n=-\infty}^{+\infty} \frac1{4\pi^2T^2n^2+4\alpha\pi Tn+\beta} \;.
\end{equation}
We can rewrite this sum as a contour integral
\begin{equation}
\frac1{2\pi i} \oint \frac{\pi\cot\pi z}{4\pi^2T^2z^2+4\alpha\pi Tz+\beta}dz\;,
\end{equation}
where the contour contains all the poles of the cotangent. The residue theorem ensures that the integral will evaluate to the sum \eqref{eindigetvoorbeeldsom}. Now we can deform the contour and turn it inside-out, which will result in it containing all poles \emph{except} for the ones of the cotangent:
\begin{equation}
- \sum_{z_0} \Res[z=z_0] \frac{\pi\cot\pi z}{4\pi^2T^2z^2+4\alpha\pi Tz+\beta} \ .
\end{equation}
The sum is now over the zeros of the polynomial in the denominator. Evaluating the residues leads to
\begin{equation}
-\frac{\cot\frac{\alpha+i\sqrt{\beta-\alpha^2}}{2T}-\cot\frac{\alpha-i\sqrt{\beta-\alpha^2}}{2T}}{4i T\sqrt{\beta-\alpha^2}} = \frac{\sinh\frac{\sqrt{\beta-\alpha^2}}T}{4T\sqrt{\beta-\alpha^2} \left( \sin^2\frac\alpha{2T} + \sinh^2\frac{\sqrt{\beta-\alpha^2}}{2T} \right)} \ .
\end{equation}
It can easily be verified that this has the correct zero-temperature limit. In exactly the same fashion one also finds that
\begin{equation}
	\sum_{n=-\infty}^{+\infty} \frac{4\pi Tn}{4\pi^2T^2n^2+4\alpha\pi Tn+\beta} = \frac{-\alpha\sinh\frac{\sqrt{\beta^2-\alpha^2}}T + \sqrt{\beta^2-\alpha^2} \sin\frac\alpha T}{2T\sqrt{\beta-\alpha^2} \left( \sin^2\frac\alpha{2T} + \sinh^2\frac{\sqrt{\beta-\alpha^2}}{2T} \right)} \;.
\end{equation}
From these equations, one can furthermore deduce that
\begin{equation}
	\sum_{n=-\infty}^{+\infty} \ln(4\pi^2T^2n^2+4\alpha\pi Tn+\beta) = \ln4\left(\sin^2\frac\alpha{2T} + \sinh^2\frac{\sqrt{\beta-\alpha^2}}{2T} \right)\;,
\end{equation}
as this is the only expression having correct $\alpha$ and $\beta$ derivatives in addition to having the right zero temperature limit.

The most useful formula is found by separating out the zero-temperature limit, which yields
\begin{equation} \label{finitetformula}
	T \sum_{n=-\infty}^{+\infty} \ln(4\pi^2T^2n^2+4\alpha\pi Tn+\beta) = \int\frac{dk_0}{2\pi} \ln(k_0^2+2\alpha k_0+\beta) + T\ln\left(1 - 2e^{-\frac{\sqrt{\beta-\alpha^2}}T} \cos\tfrac\alpha T + e^{-2\frac{\sqrt{\beta-\alpha^2}}T} \right) \;.
\end{equation}

\bibliography{biblio}

\end{document}